\documentclass[prl,showpacs,superscriptaddress,twocolumn,floatfix,amsmath,amssymb]{revtex4-1}%
\usepackage[dvips]{graphicx}
\usepackage{xspace}
\usepackage{color}
\usepackage[dvipsnames]{xcolor}
\usepackage{array}
\usepackage{hyperref}
\usepackage[capitalize]{cleveref}
\usepackage{multirow}
\usepackage{braket}
\usepackage{tabularx}

\usepackage{amstext} %
\usepackage{array}   %
\newcolumntype{L}{>{$}l<{$}} %
\newcolumntype{C}{>{$}c<{$}} %

\newcommand{\rucl}{$\alpha\mathrm{-RuCl}_3$\xspace}
\newcommand{\cstar}{\ensuremath{{c^{\ast}}}}

\newcommand{\eps}{\ensuremath{\epsilon_\cstar}}
\renewcommand{\eps}{\ensuremath{\epsilon}}
\newcommand{\compressibility}{\kappa_{\cstar \cstar}}
\newcommand{\heldconstant}{\ensuremath{X}}

\newcommand{\optional}[1]{\textcolor{gray}{#1}}
\renewcommand{\optional}[1]{}

\newcommand{\MEC}[1]{\ensuremath{\widetilde{#1}}}

\newcommand{\Gruneisen}{\ensuremath{\Gamma_\text{s}}}

\newcommand{\Jij}{\ensuremath{\mathbb{J}_n^{\gamma}}}

\newcommand{\itsection}[1]{\textit{#1.}---}

\newcommand{\supplcitations}{\nocite{suzuki2020quantifying,johnson2015monoclinic,johnson2015monoclinic,johnson2015monoclinic,winte18,leahy2017anomalous,modic2018resonant,modic2018chiral,modic2018resonant,modic2020scale,riedl2019sawtooth,gass2020field,winte17,winte17,winte17,maksimov2020rethink,bachus2020thermodynamic,winte17,bachus2020thermodynamic,bachus2020thermodynamic,bachus2020thermodynamic,winte17,bachus2020thermodynamic,bachus2020thermodynamic,gass2020field,winte17,winte17,qe,pbe,grimme,mp,vasp,blochl,kim2016crystal,hermann2018competition,hermann2019pressure,johnson2015monoclinic,banerjee2017neutron,do2017majorana,lampen2018field,slater1960quantum,eichstaedt2019deriving,eichstaedt2019deriving,montalti2006handbook,jackeli2009mott,rau2014trigonal,winter2017models,chaloupka2013zigzag,foyevtsova2013abinitio,kim2015kitaev,winte16,winte16,hou2017unveiling,neese2012orca,rau2016spin,winter2017models,laurell2020dynamical}}

\begin{document}
\newcommand{\frankfurt}{Institut f\"ur Theoretische Physik, Goethe-Universit\"at Frankfurt,
Max-von-Laue-Strasse 1, 60438 Frankfurt am Main, Germany}
\newcommand{\salem}{Department of Physics and Center for Functional Materials, Wake Forest University, Winston-Salem, North Carolina 27109, USA}

\author{David A. S. Kaib}
\email[]{kaib@itp.uni-frankfurt.de}
\affiliation{\frankfurt}
\author{Sananda Biswas} %
\affiliation{\frankfurt}
\author{Kira Riedl}
\affiliation{\frankfurt}
\author{Stephen M. Winter}
\affiliation{\frankfurt}
\affiliation{\salem}
\author{Roser Valent{\'\i}}
\affiliation{\frankfurt}
\date{\today}
\title{Magnetoelastic coupling and effects of uniaxial strain in $\alpha$-RuCl$_3$ from first principles}

\begin{abstract}
We present first-principles results on the magnetoelastic coupling in \rucl and uncover
	a striking dependence of the magnetic coupling constants on strain effects. %
	Different magnetic interactions are found to respond very unequally to variations in the lattice, with the Kitaev interaction being the most sensitive. 
	Exact diagonalization results on our magnetoelastic model reproduce recent measurements of the structural Gr\"uneisen parameter and explain the origin of the negative magneto\-striction of \rucl, disentangling contributions related to different anisotropic interactions and $g$ factors. 
	Uniaxial strain perpendicular to the honeycomb planes %
	is predicted to reorganize the relative coupling strengths, strongly enhancing the Kitaev interaction while simultaneously weakening the other anisotropic exchanges under compression. Uniaxial strain may therefore pose a fruitful route to experimentally tune \rucl nearer to the Kitaev limit. 
\end{abstract}

\maketitle

\renewcommand{\Jij}{\mathbb J_{ij}}

The exactly solvable Kitaev honeycomb model \cite{kitaev2006anyons} features a quantum spin liquid (QSL) with non-Abelian anyons under magnetic fields. 
Following the proposal to realize the highly frustrated Kitaev interaction in real materials through an intricate exchange mechanism  \cite{jackeli2009mott}, so-called ``Kitaev-candidate materials'' emerged \cite{winter2017models,hermanns2018physics,takagi2019concept}. 
These are spin-orbit Mott insulators, whose low-energy magnetic degrees of freedom can be described through $j_\text{eff}=1/2$ \textit{pseudospins}.
So far, most candidate materials exhibit long-range ordered magnetic ground states  \cite{liu2011long,ye2012direct,biffin2014unconventional,sears2015magnetic,johnson2015monoclinic,williams2016incommensurate} instead of the Kitaev QSL due to residual extended interactions beyond the pure Kitaev model \cite{chaloupka2010kitaev,chaloupka2013zigzag,rau2014generic,biffin2014noncoplanar,lee2015theory,williams2016incommensurate}. 
Nevertheless, the physics of such extended Kitaev models have lead to 
countless interesting unconventional phenomena in these materials
with arguably the most prominent example being \rucl. 
With the goal of tuning away from its antiferromagnetic zigzag order and possibly to a Kitaev QSL, various routes have been considered, including chemical doping \cite{koitzsch2017nearest,bastien2019spin,baek2020}, 
graphene substrates \cite{mashhadi2019spin,zhou2019evidence,biswas2019electronic,gerber2020ab}, %
 hydrostatic pressure \cite{biesner2018detuning,bastien2018pressure,wang2018pressure,yadav2018strain,li2019raman}
 and magnetic fields %
 \cite{wolter2017field,baek2017evidence,wang2017magnetic,banerjee2018excitations,kasahara2018majorana,balz2019finite,yokoi2020half,yamashita2020sample}. 
 In the case of hydrostatic pressure, dimerization quickly destroys the $j_\text{eff}=1/2$ picture \cite{biesner2018detuning,li2019raman} such that no Kitaev QSL can occur.
\rucl under magnetic fields has however attracted great attention, due to the observation of a narrow field-induced regime of quantized thermal Hall conductivity \optional{in some samples}  \cite{kasahara2018majorana,yokoi2020half,yamashita2020sample}. 
Subsequent theoretical studies  highlighted the importance of magnetoelastic coupling for the description of the thermal Hall conductivity \cite{vinkler2018approximately,ye2018quantization} and investigated further consequences of magnetoelastic coupling \cite{metavitsiadis2020phonon,ye2020phonon}, in both cases for the idealized pure Kitaev model.
The behavior of the longitudinal thermal conductivity under magnetic field already
implies a strongly magnetoelastically-coupled phonon heat transport \cite{hentrich2018UnusualPhononHeat,hentrich2020HighfieldThermalTransport}. 
Therefore realistic microscopic modeling of magnetoelastic coupling, taking the actual lattice and the extended (non-Kitaev) interactions into account, is crucial in tackling this key issue of \rucl. 
In contrast to conventional spin-lattice coupling, both the spin-orbital nature of the pseudospins \cite{liu2019JahnTeller,porras2019pseudospin} and the geometry-sensitive exchange mechanisms of Kitaev materials \cite{jackeli2009mott,rau2016spin,winter2017models} 
indicate 
\emph{pseudospin}-lattice coupling to be more delicate. 

 \begin{figure}
\includegraphics[width=\linewidth]{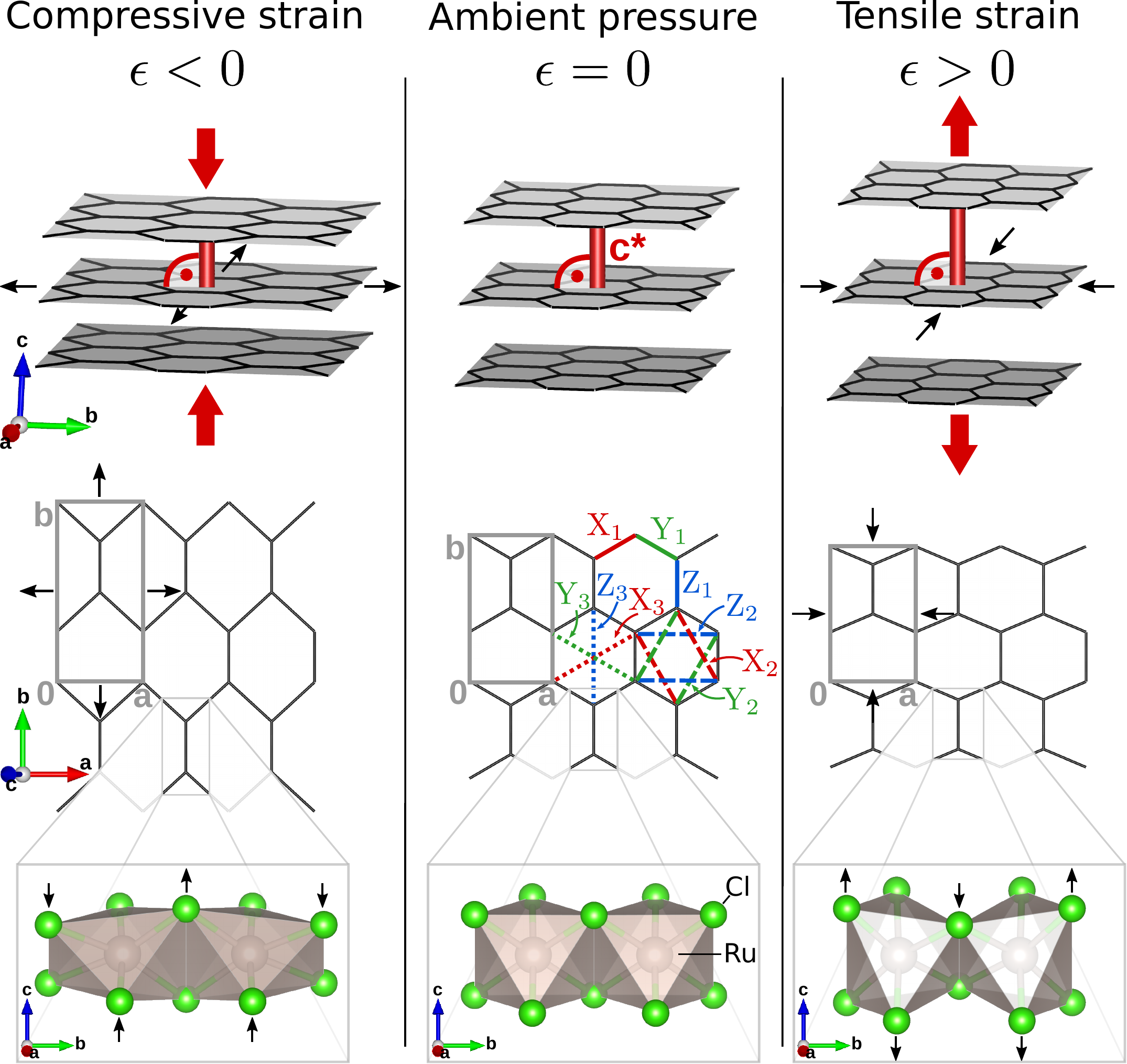}
  \caption{%
   Structural effects as a consequence of compressive [left column] and tensile [right column] uniaxial strain onto~\cstar
   . The middle column shows the unstrained structure. 
    Enforced strains are indicated by red arrows and predicted responses of the system by black arrows. %
Shown from top to bottom are:
Honeycomb layers, view onto one layer, and Ru-Ru bond with  chlorine
	 environment. 
  \label{fig:schematic} 
  }
\end{figure}

In this Letter we explore how the extended interactions in \rucl are coupled to \emph{uniaxial} strain. Here we focus on strain perpendicular to the honeycomb planes (parallel to \cstar, see  \cref{fig:schematic}), as it should be the direction easiest to tune experimentally and as its magnetostriction has been measured recently \cite{gass2020field}. 
 By combining first-principles simulations and exact diagonalization we unveil a subtle dependence of the magnetic coupling constants on strain effects and provide a microscopic understanding of magnetoelastic properties in \rucl.

\itsection{Magnetoelastic model}
We first derive
 the magneto\-elastic Hamiltonian of \rucl\  
under a magnetic field $\mathbf B$ with uniaxial strain $\eps\equiv \Delta\cstar/c^\ast_{\,0}$
as a degree of freedom,
\begin{equation} \label{eq:Hamil}
\mathcal{H}
= 
\sum_{ij}
\mathbf{S}_i \cdot \Jij(\eps) \cdot  \mathbf{S}_j - \mu_B \sum_i \mathbf{B} \cdot \mathbb{G}(\eps) \cdot \mathbf{S}_i
.
\end{equation}
$\cstar$ is the distance between the honeycomb layers [\cref{fig:schematic}] and $\mathbf S_i$ are $j_{\text{eff}}=1/2$ operators. The strain-dependent tensors $\Jij(\eps)$ and $\mathbb G(\eps)$ contain all exchange and $g$-tensor couplings.  
Our primary objective is then to extract the strengths of the linear magnetoelastic couplings 
$\MEC{\mathcal J}\equiv \left.\left(\frac{\partial \mathcal{J} }
{\partial \eps}
\right)
\right|_{\eps=0} 
$
for all components $\mathcal J\in \Jij$ and $\mathcal J \in \mathbb G$.
This way we (i)~explore uniaxial strain as a potential tuning parameter in future experiments and (ii)~enable theoretical modeling of observables that directly couple magnetic and structural degrees of freedom.  
We then apply our obtained magnetoelastic model to the field-dependent structural Gr\"uneisen parameter and magnetostriction, finding good agreement with recent measurements~\cite{gass2020field}.

For Kitaev materials the interaction tensor $\Jij$ %
is highly
anisotropic and bond-directional-dependent.  Bond types called X$_n$,Y$_n$ and Z$_n$ are defined on $n$th-nearest neighbors as shown in the center of \cref{fig:schematic}. The exchange is then 
\begin{equation}
\Jij = \left(
 \begin{array}{c|ccc} 
 &\alpha&\beta&\gamma\\
 \hline
 \alpha & J_n & \Gamma_n + D^\gamma_n & \Gamma_n^\prime - D^{\beta}_n \\
 \beta & \Gamma_n - D^{\gamma}_n& J_n & \Gamma_n^\prime + D^\alpha_n \\
 \gamma & \Gamma_n^\prime + D^\beta_n & \Gamma_n^\prime - D^\alpha_n& J_n+K_n
 \end{array}\right)
 ,
\end{equation}
where $(\alpha,\beta,\gamma) = (x,y,z)$ for Z$_n$-bonds, $(y,z,x)$ for X$_n$, and $(z,x,y)$ for Y$_n$-bonds. When $K_1$ is the only finite coupling, the model reduces to the exactly solvable Kitaev model \cite{kitaev2006anyons}. 
The Dzyaloshinskii-Moriya interaction $(D
^\alpha_n, D^\beta_n, D^\gamma_n)$ vanishes for $n=1,3$ due to inversion symmetry. 
For simplicity we employ $C_3$-symmetrized models throughout this manuscript, such that coupling strengths on \mbox{X$_n$-,} Y$_n$- and Z$_n$-bonds are equal for a given $n$. 
Deviations from this $C_3$-symmetry within the C2$/m$ space group are discussed in  Supplemental Material (SM)~\cite{Suppl}\supplcitations.

\itsection{First-principles methods}
To include effects beyond a homogeneous elongation of the lattice with $\cstar$-strain, we employ constrained geometric optimizations. 
To obtain a zero-strain starting structure, the ambient-pressure experimental C2/$m$ structure \cite{cao2016low} was fully relaxed, including all lattice parameters and internal atomic positions. Subsequently, %
the lattice parameters $a$, $b$, monoclinic angle $\beta$, and atomic positions were relaxed while constraining $c$ to different values.  
For each obtained structure, the strain is then $\eps=\Delta \cstar/ c^\ast_{\,0}$, with $\cstar=c \sin\beta$ and $c^\ast_{\,0}$ 
denoting the unstrained parameter. 
The constrained relaxations were performed within GGA+$U$~\cite{pbe,Suppl} in 
zigzag antiferromagnetic configurations using %
Quantum Espresso~\cite{qe}.

To determine the strain-dependent $g$-tensor components, we computed $\mathbb G$ for each relaxed geometry on
[RuCl$_6$]$^{3-}$ molecules with the quantum chemistry ORCA 3.03
package~\cite{neese2012orca,neese2005efficient} 
with the functional TPSSh, basis set def2-TZVP, and complete active space for the $d$ orbitals CAS(5,5) ---
 an approach that has proved reliable for isolated $d^5$ molecules
\cite{pedersen2016iridates}. 

For the exchange interactions $\Jij(\eps)$, we first
computed non-relativistic hopping parameters for each relaxed structure in
non-spin-polarized configurations within GGA using the Full Potential Local Orbital
(FPLO) code \cite{fplo}. 
Magnetic interactions were then estimated via exact diagonalization of the two-site five-orbital Hubbard Hamiltonian
 and projection of the low-energy states onto the $j_{\text{eff}}=1/2$ subspace \cite{riedl2019abinitio,winte16}. 
Here, we considered both $t_{2g}$ and $e_g$
orbitals explicitly, extending on previous approaches of some of the
authors \cite{winte16}.  
 Further details on first-principles calculations are given in 
SM~\cite{Suppl}.

\itsection{First-principles results} 
The predicted  effects of compressive (negative) and tensile (positive) uniaxial strain on the structure are summarized 
 in \cref{fig:schematic} (showing illustrative extreme strains). 
  \begin{figure}
	\includegraphics[width=\linewidth]{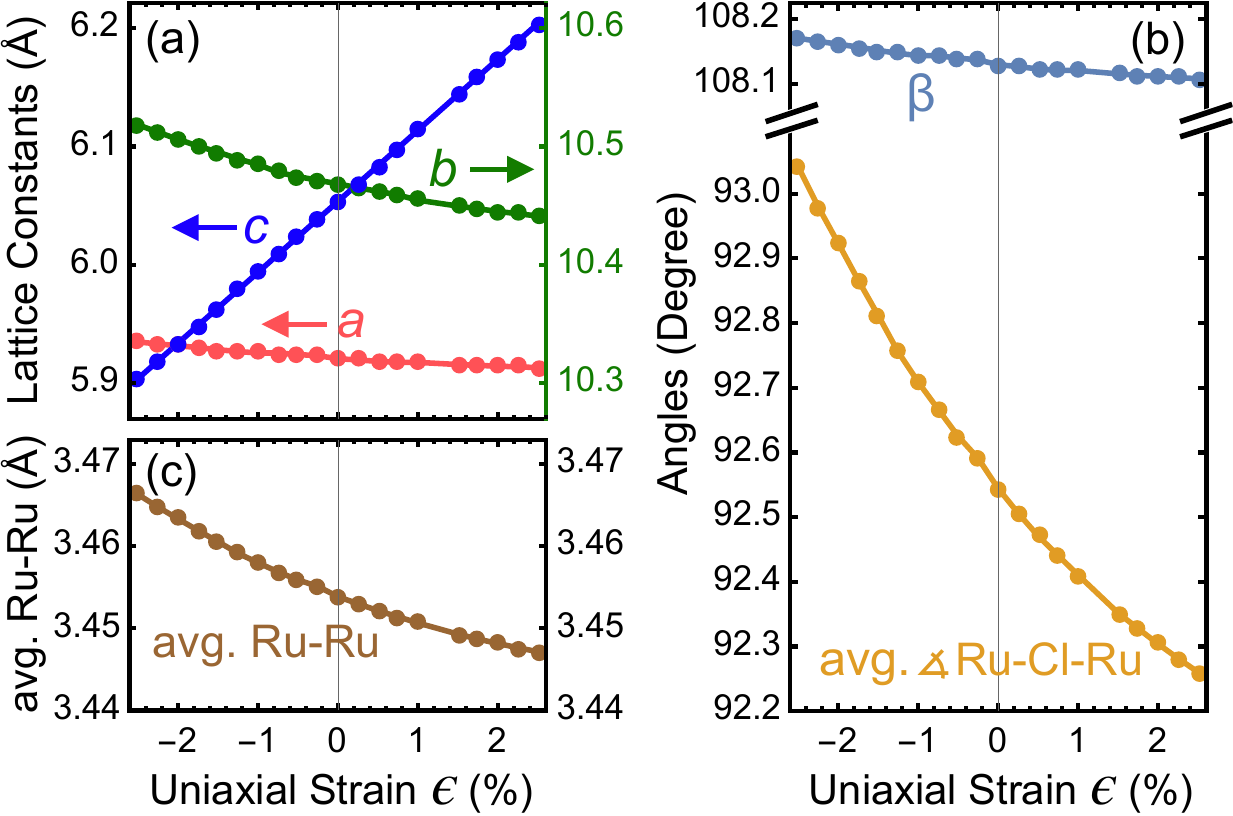}
	\caption{
	Relaxed lattice parameters as a function of uniaxial strain on $\cstar$. (a)~Lattice constants. $b$ is to be read with the right axis. (b)~Monoclinic angle $\beta$ and angle of the Ru-Cl-Ru bonds (average over X$_1$, Y$_1$, Z$_1$ bonds).  (c)~Average Ru-Ru bond length. 
	\label{fig:strain_vs_lattice}}
\end{figure}
Quantitative results are shown in Fig.~\ref{fig:strain_vs_lattice}. 
 Upon compression along $\cstar$, the honeycomb $ab$ plane expands, increasing the Ru-Ru distance. %
 {Importantly, the octahedral chlorine environment, whose precise geometry mainly governs the Jackeli-Khaliulin exchange mechanism \cite{jackeli2009mott,rau2014trigonal,winter2017models}, is distorted in a strongly non-homogeneous way under uniaxial strain, see bottom row of \cref{fig:schematic}.} %

\begin{table*}
\begin{tabular}{L||C|C||C|C|C|C||C|C|C|C|C|C|C||C|C|C|C}
\hline 
 &g_{ab}  & g_\cstar & J_1 & K_1 & \Gamma_1 & \Gamma'_1 & J_2 & K_2 & \Gamma_2 & \Gamma'_2 & D^\alpha_{2} & D^\beta_{2}& D^\gamma_{2} & J_3 & K_3 & \Gamma_3 & \Gamma'_3  \\
  \hline 
 \mathcal J|_{\eps=0}%
  & 2.36 & {1.88} & \mathbf{-5.7} & \mathbf{-10.1} &  \ \mathbf{9.3}\, & \mathbf{-0.7} & 0. & -0.2 & 0.1 & 0. & 0. & 0. & 0.1 & \ 0.2\, & \ 0.2\, & 0. & -0.1 \\
\MEC{\mathcal J} & {-1.6} & {3.85} & 1.3 & \mathbf{40.5} & \ \mathbf{7.5}\, & \mathbf{-11.5} & -0.9 & 1.6 & -0.4 & -0.1 & -1. & -1. & -3.2 & \ 1.6\, & \ 0.6\, & -0.6 & -0.5 \\
\hline
\end{tabular}
\caption{$C_3$-symmetrized magnetic couplings at ambient pressure $\mathcal
	J|_{\eps=0}$ and the associated magnetoelastic couplings
	$\MEC{\mathcal J}\equiv (\partial \mathcal J/\partial \eps) |_{\eps=0}$. Except
	for unitless $g$-tensor components, units are in meV. Strongest
	couplings are highlighted. 
\label{tab:couplingssummary}
}
\end{table*}

\begin{figure}
\includegraphics[width=\linewidth]{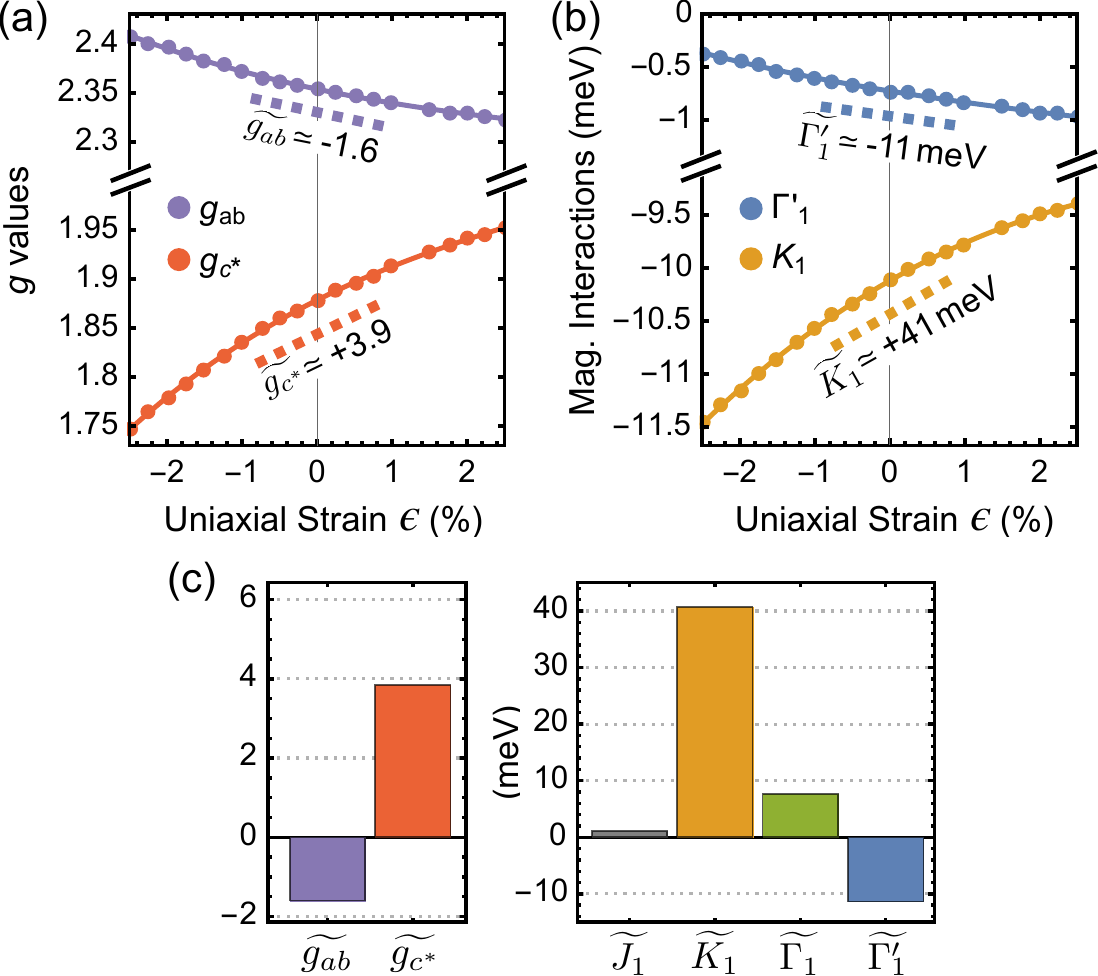}
  \caption{
	  (a) Calculated strain-dependence of $g$ values and (b) $K_1$, $\Gamma'_1$  (colored bullets). 
  Solid curves show third-order polynomial fits. 
  Dashed lines indicate $\MEC{\mathcal J}\equiv (\partial \mathcal{J}/\partial \eps) |_{\eps=0}$.
  (c)~Comparison of magnetoelastic couplings $\MEC{\mathcal J}$ for $g$ values and nearest-neighbor interactions. 
\label{fig:barplot}
}
\end{figure}

The magnetoelastic couplings $\MEC{\mathcal J}\equiv (\partial \mathcal{J}/ \partial \eps) |_{\eps=0}$ for each %
 $g$-value ($\mathcal J \in \mathbb G$) and magnetic coupling parameter ($\mathcal J \in \Jij$)
were determined by differentiating third-order polynomial fits to their strain-dependencies
as illustrated exemplary in Fig.~\ref{fig:barplot}(a,b) for the couplings with the strongest strain-dependence. %
Corresponding magneto\-elastic couplings of the $g$ values and nearest-neighbor interactions are compared in \cref{fig:barplot}(c). 
The complete set of obtained ambient-pressure model parameters $\mathcal J|_{\eps=0}$ and $\MEC{\mathcal J}$ is listed in Table~\ref{tab:couplingssummary}, with large couplings highlighted.

The gyromagnetic tensor $\mathbb G$ of each magnetic site is determined mainly by its local chlorine environment.
Due to the non-trivial distortion under uniaxial strain~\eps\ [\cref{fig:schematic}],
the strain-dependence of the $g$-anisotropy cannot be explained with regular expressions~\cite{chaloupka2016magnetic} that are valid for trigonally compressed octahedral environments. 
From \textit{ab-initio}, the non-negligible components of $\mathbb G$ for the zero-strain structure are found to be:
$g_{ab} = 2.36$ (in-plane) and $g_\cstar=1.88$ (out-of-plane),
which fall in the range of existing estimates~\cite{yadav2016kitaev,chaloupka2016magnetic,winte18,sahasrabudhe2020high}. %
For  their magnetoelastic couplings, we extract $\MEC{g_{ab}} = -1.6$ and $\MEC{g_\cstar}=3.85$. 
Compressive strain $\eps<0$ will therefore increase $g_{ab}$ and decrease $g_\cstar$, enhancing the $g$ anisotropy further, see \cref{fig:barplot}(a).

The magnetic interactions $\Jij$ are mainly governed by the corresponding Ru-Ru distances and the Ru-Cl-Ru geometry [Fig.~\ref{fig:strain_vs_lattice}(b,c)] 
through modified orbital overlap integrals. 
Inclusion of virtual processes involving the $e_g$ orbitals %
is found to strongly renormalize some interactions \cite{Suppl}, doubling, for example, the magnitude of the Kitaev exchange $K_1$. 
This interaction constitutes the strongest coupling in the ambient-strain model ($K_1=-10.1\,\text{meV}$) and has the strongest strain dependence ($\MEC{K_1}=40.5\,\text{meV}$), see \cref{fig:barplot}(c). 
The fact that %
the large $\MEC{K_1}$ has opposite sign of $K_1$ implies that \emph{compressive} strain ($\eps<0$) firmly \emph{strengthens} the Kitaev interaction. %
Regarding the results on the set of extended interactions as a whole, we emphasize that uniaxial strain  %
affects different interactions
unequally, 
 strengthening some while weakening others --- in contrast to effects predicted for volumetric strain or hydrostatic pressure \cite{yadav2018strain}.  %
 Inspecting again the effect of compressive \cstar-strain, the 
shared sign of $\MEC{\Gamma_1}$ with $\Gamma_1$ implies that $|\Gamma_1|$ will be \emph{weakened}, which analogously holds for $|\Gamma'_1|$. The structure of the largest magnetoelastic couplings [bold in second row of \cref{tab:couplingssummary}] therefore implies that compressive \cstar-strain predominantly shifts interaction strength away from these  anisotropic couplings and towards the Kitaev exchange $K_1$.

Contrary to what one may expect, %
$\cstar$-strain predominantly couples to these in-plane interactions, %
whereas \mbox{\emph{inter}}-plane magnetoelastic couplings are found to be much weaker \cite{Suppl}. 
Experiments probing $\cstar$-variations are therefore highly sensitive to the in-plane magnetism.

\itsection{Discussion}
For the application of our derived models we first focus on primarily magnetic observables.  These are determined mainly by the zero-strain interactions $\mathcal J|_{\eps=0}$ [first row in \cref{tab:couplingssummary}] and can be computed using ED in the projected $j_\text{eff}=1/2$ basis on a hexagon-shaped 24-site cluster. %
 Throughout, we find very good %
 agreement with experimental observations. In particular, the zigzag-ordered ground
 state%
 , correct critical field strengths \cite{baek2017evidence,wolter2017field} and the evolution of the magnetic torque \cite{leahy2017anomalous,modic2018resonant,modic2018chiral} and magnetotropic coefficient \cite{modic2018resonant,modic2020scale} are captured. %
Peculiarly, a ferromagnetic phase is highly proximate to the ground state, and zigzag
order is only upheld by the weak $\Gamma'_1=-0.7\,\text{meV}$. 
The large $\MEC{\Gamma'_1}=-11.5\,$meV therefore implies that compressive $\cstar$-strain
 should strongly destabilize zigzag order. 
Detailed results for magnetic properties are shown in SM \cite{Suppl}.

Our main focus lies on \emph{magnetoelastic} properties, which are driven by $\MEC{\mathcal J}$. 
 Motivated by recent measurements by Gass \textit{et al}.~\cite{gass2020field}, we focus on
 linear magnetostriction $\lambda_\cstar \equiv \cstar^{-1} (\partial \cstar/\partial B)
 $ and the %
 structural Gr\"uneisen
 parameter $\Gruneisen \equiv -\frac{(\partial S_\text{m}/\partial
 p_\cstar)}{T(\partial S_\text{m}/\partial T)}$.
 Here $p_\cstar$ is uniaxial pressure along $\cstar$ and  $S_\text{m}$ the magnetic entropy, which the authors of Ref.~\onlinecite{gass2020field} obtained via subtraction of phononic contributions. 
Under the assumption that the diagonal components of the elasticity tensor are dominant,
the observables can be approximated~\cite{Suppl}
\begin{align}
  \lambda_\cstar 
  &\approx\frac{\compressibility}{V} \sum_{{\mathcal J}
  \in{\Jij,\mathbb G}
  }
  \, \MEC{\mathcal J} \,
  \left( \frac{\partial M}{\partial {\mathcal J}} \right)_{\eps=0}
  ,
  \label{eq:alambdafin}
\\    \Gruneisen 
   &\approx  \frac{\compressibility}{T} \sum_{{\mathcal J} 
     \in{\Jij,\mathbb G}}
 \,  \MEC{\mathcal J} \,
   \left( \frac{\partial S_\text m}{\partial \mathcal J}\right)_{\eps=0} \left(\frac{\partial S_\text m}{\partial T}\right)^{-1}_{\eps=0} , \label{eq:aGruneisenfin}
\end{align}
 where the sums go over all strain-dependent interactions and $g$ values:
 $\mathcal J\in\{J_1,K_1,\dots,g_{ab},g_\cstar\}$.  The magneto\-elastic couplings $\MEC{\mathcal J}$ are taken from \cref{tab:couplingssummary} and the derivatives are evaluated at $\eps=0$ (\textit{i.e.}, at parameters $\left.\mathcal J\right|_{\eps=0}$) within ED. 
 We compute quantities up to the unknown $\compressibility \equiv -(\partial \eps / \partial
 p_\cstar)$ of \rucl, defined as
 the linear compressibility along $\cstar$ against
 \textit{uniaxial} pressure~$p_\cstar$.

\begin{figure}
\includegraphics[width=1\linewidth]{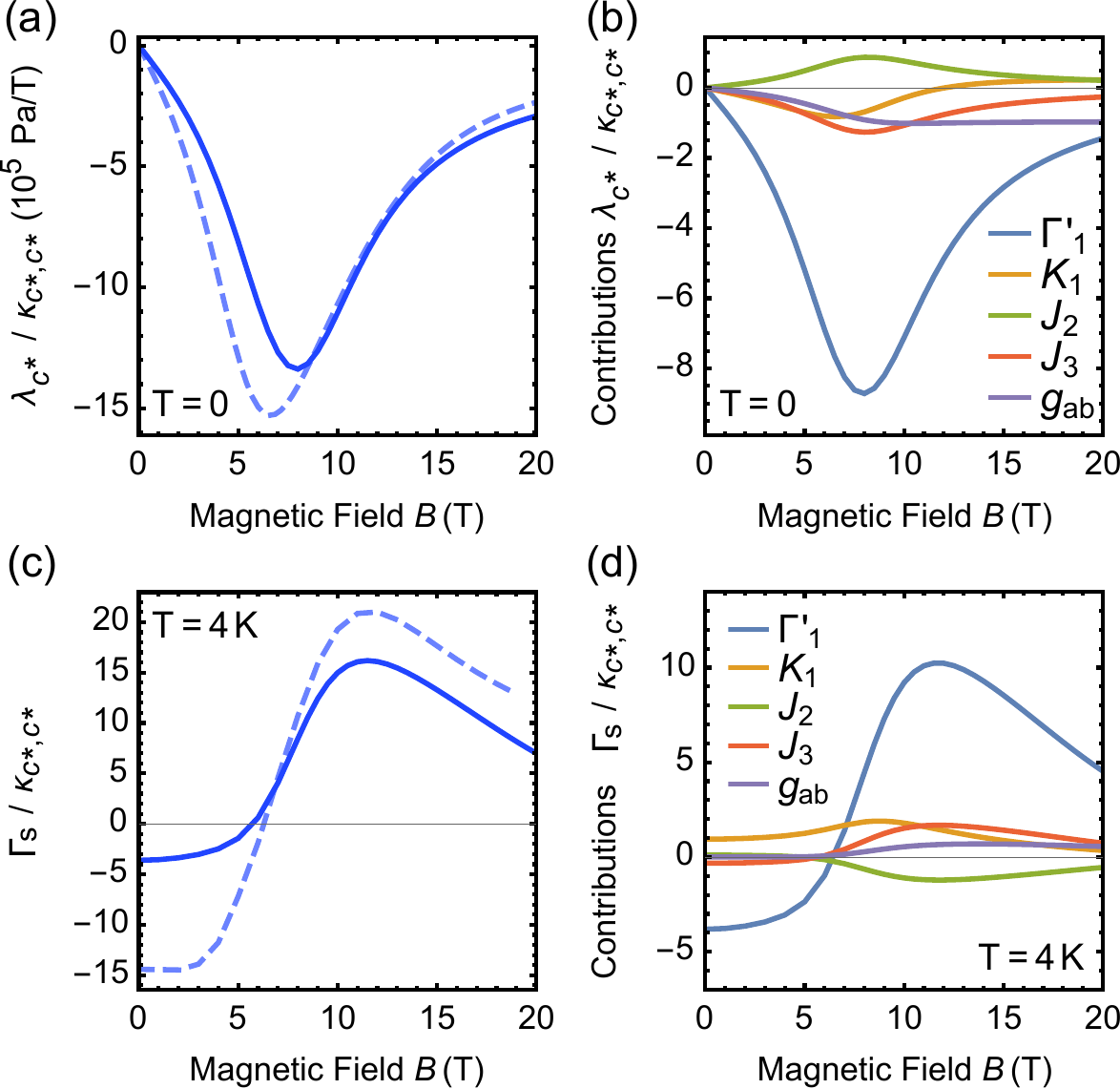}
  \caption{
	  (a) Calculated field-dependent magnetostriction  %
	at $T=0$\,K and (c) structural Gr\"uneisen parameter %
 at $4$\,K. %
 $\mathbf B \parallel b$. 
  (b,d)~Dissection of the largest contributions to the respective quantities via \cref{eq:alambdafin,eq:aGruneisenfin}. 
Solid lines: Obtained using all $\MEC{\mathcal J}$ and $\mathcal J|_{\eps=0}$ from \cref{tab:couplingssummary}. Dashed lines: Using all $\MEC{\mathcal J}$ from \cref{tab:couplingssummary} and $\mathcal J|_{\eps=0}$ from the model of Ref.~\onlinecite{winte17}. 
  \label{fig:lambdaGamma}
}
\end{figure}

In \cref{fig:lambdaGamma} we present results for $\lambda_\cstar$
and $\Gruneisen$ obtained from the \cref{tab:couplingssummary} parameters as solid curves %
(dashed curves %
are discussed below).
\cref{fig:lambdaGamma}(a) shows
the $T\rightarrow 0$ magnetostriction $\lambda_\cstar$ %
as a function of in-plane field $\mathbf B\parallel b$.
The magnetostriction exhibits its maximum magnitude at the critical field of the model ($B_c\approx 7.5\,$T). Note that finite-size effects in ED typically broaden features near the critical field, hence $\lambda_\cstar$ is expected to peak sharper at $B_c$ in the thermodynamic limit \cite{Suppl}. For increasing field strengths $B>B_c$, the magnitude of $\lambda_\cstar$ shrinks monotonically. The  \emph{negative} magnetostriction throughout implies a field-induced compression of \rucl.  
We therefore find very good agreement with  experiment \cite{gass2020field}, although it is not clear whether a subtle reported kink above $B_c$ \cite{gass2020field} is also present in our results. %

The form of \cref{eq:alambdafin} allows to dissect individual contributions to the magnetostriction, $\MEC{\mathcal J}\cdot (\partial M/\partial \mathcal J)$,  stemming from the interplay of different interactions $\mathcal J$ with the lattice.  %
One may expect that the magnetostriction would be influenced strongly by the summand with $\mathcal J=K_1$ due to the  large $\MEC{K_1}$ (cf.~\cref{fig:barplot}(c)). %
However, we find the associated
magnetization susceptibility $(\partial M/\partial K_1)$ to be negligible in
magnitude compared to other magnetization susceptibilities. Instead, the magnetostriction is found to be governed by the summand with  $\mathcal J=\Gamma'_1$, as shown in \cref{fig:lambdaGamma}(b).  Here
 $(\partial M/\partial \Gamma'_1)>0$, which can already
be anticipated on the classical level \cite{Suppl}, and $\MEC{\Gamma'_1}<0$. The negative
magnetostriction may therefore be understood as follows: Under increased $B$,
\rucl\ can lower its Zeeman energy further (\textit{i.e.}, increase its magnetization)
 by
\emph{increasing} $\Gamma'_1$ ($\partial M/\partial \Gamma'_1>0$), which is
achieved by \cstar-\emph{compression} ($\MEC{\Gamma'_1}<0$).

We now turn to the structural Gr\"uneisen parameter \Gruneisen, computed via \cref{eq:aGruneisenfin} and ED. Here, we achieve finite temperatures by restricting the canonical sums to the lowest 16 eigenstates, %
which works well for lowest temperatures \cite{Suppl}. 
Results are shown in \cref{fig:lambdaGamma}(c,d). We again find a good qualitative agreement with experiment \cite{gass2020field}, with a sign change from negative to positive near $B=B_c$. 
Likely because of finite-size effects, the slope at the sign change is not vertical, and $\Gruneisen$ only reaches its maximum for fields slightly above $B_c$. 
In contrast to experiment, we obtain
$|\Gruneisen(B\lesssim B_c)|<|\Gruneisen(B\gtrsim B_c)|$. 
Dissecting the contributions from different magnetoelastic couplings in \cref{fig:lambdaGamma}(d), we again find that the contribution related to $\MEC{\Gamma'_1}$ dominates the magnetoelastic response. Analyzing our results for temperatures below that of the experiment (4\,K), we predict most of the qualitative response to be unchanged. However, we note that our model also predicts an anomalous drop in both the structural (\Gruneisen ) and magnetic ($\Gamma_{B}$) Gr\"uneisen parameters at high fields $B\approx 22$\,T \cite{Suppl,bachus2020thermodynamic}, which becomes increasingly sharp at lower temperatures. Experimentally, such an anomaly was clearly observed in $\Gamma_{B}$ around $B\approx 10\,$T at $T < 2$\,K \cite{bachus2020thermodynamic}, and is also suggested by recent data on \Gruneisen\ at $T=3.5$\,K in the same field range \cite{gass2020field}. These anomalies are understood to occur due to interchange of lowest excited states \cite{bachus2020thermodynamic}, which occurs at a field strength that is highly sensitive to the specific couplings. This may be considered for future refinements of the model.

 In the results discussed so far, we employed the complete set of magnetic interactions $\mathcal J|_{\eps=0}$ and magnetoelastic couplings $\MEC{\mathcal J}$ from \cref{tab:couplingssummary}. 
However, our results on the magnetoelastic coupling may also provide guidelines for theoretical modeling in reduced parameter spaces. 
Therefore we also considered a \emph{minimal} magnetic model $\{\mathcal J|_{\eps=0}\}$ with only four nonzero interactions, $(J_1,K_1,\Gamma_1,J_3)=(-0.5,-5.0,2.5,0.5)\,$meV and $g_{ab}=2.3$, which has reproduced key experimental observations on magnetic properties \cite{winte17,wolter2017field,cookmeyer2018spin,winte18,riedl2019sawtooth,sahasrabudhe2020high,bachus2020thermodynamic}. 
We repeated our %
calculations, %
evaluating the derivatives in \cref{eq:alambdafin,eq:aGruneisenfin} at these minimal-model values $\mathcal J|_{\eps=0}$  (keeping the magnetoelastic couplings $\MEC{\mathcal J}$ from  \cref{tab:couplingssummary}). 
Results are shown as dashed lines in \cref{fig:lambdaGamma}(a,c). Overall, the results are comparable to before. 
Note ---importantly--- that a summand in \cref{eq:alambdafin,eq:aGruneisenfin} related to $\MEC{\mathcal J}$ is \emph{not} required to vanish 
 if the respective $\mathcal J|_{\eps=0}$ is zero. 
On the contrary, we find also in the case of this minimal  model that contributions from the magneto\-elastic coupling  $\MEC{\Gamma'_1}$ essentially dominate the response, regardless of  $\left.\Gamma'_1\right|_{\eps=0}=0$. This highlights that the dominant magnetoelastic interactions in \rucl\ can be of completely different form than the dominant ambient-pressure magnetic interactions.

\itsection{Conclusions}
We derived a magnetoelastic Hamiltonian for \rucl\ completely from  \textit{ab-initio}. We have shown that it reproduces key magnetic phenomena of \mbox{\rucl}\ and can explain recent field-dependent structural Gr\"uneisen and magnetostriction measurements \cite{gass2020field}. For magnetoelastic properties, a $\Gamma'_1$-type magnetoelastic coupling (``$\MEC{\Gamma'_1}$'') is found to dominate, albeit the associated interaction $\Gamma'_1$ being subdominant in the purely magnetic part of the Hamiltonian. 
Such non-Kitaev magnetoelastic effects should be reconsidered when comparing  pure-Kitaev pseudospin-phonon modeling 
with experiments. 
Compressive uniaxial strain  perpendicular to the honeycomb planes is predicted to strongly destabilize zigzag order while shifting interaction strength from other anisotropic couplings towards the Kitaev exchange. 
The strong reorganization of the magnetic interactions by uniaxial strain is a result of the geometry-sensitive exchange mechanisms in Kitaev materials and therefore likely extends also to other two- and three-dimensional Kitaev materials. 
The methodology we explored in this study is extendable to other materials and arbitrary strain fields. It enables to tackle quantitatively the pseudospin-phonon couplings, 
that may play a crucial role in understanding the thermal Hall conductivity. 

\textit{Note added:}
After completion of this work, several experimental studies were posted that further highlight the importance of magnetoelastic coupling in \rucl \cite{schoenemann2020thermal,hentrich2020HighfieldThermalTransport,li2020fractional}. 

\begin{acknowledgments}
\itsection{Acknowledgments}
We thank Anja Wolter and Bernd B\"uchner for fruitful discussions
and we acknowlegde the Deutsche Forschungsgemeinschaft
(DFG, German Research Foundation) for
funding through Project No.\ 411289067 (VA117/15-1)
	and TRR 288 - 422213477 (project A05).
\end{acknowledgments}

\bibliography{Notes.bib}

\clearpage
\widetext
\appendix
\begin{center}
\textbf{\large \textit{Supplemental Material}:\\ \smallskip Magnetoelastic coupling and effects of uniaxial strain in \rucl\ from first principles} \bigskip \bigskip
\end{center}
\twocolumngrid

\setcounter{equation}{0}
\setcounter{figure}{0}
\setcounter{table}{0}
\setcounter{page}{1}
\makeatletter
\renewcommand{\theequation}{S\arabic{equation}}
\renewcommand{\thefigure}{S\arabic{figure}}
\renewcommand{\thetable}{S\Roman{table}}

\section{Magnetic properties of the \textit{ab-initio} derived model}
We study primarily magnetic properties of the $\{\mathcal J|_{\eps=0}\}$ model given in Table~I of the main text. While the physics of these properties have been covered and discussed in previous modeling, they provide comparisons to a wide array of measurements. Therefore the results presented in this section serve primarily as benchmark of our fully \textit{ab-initio} obtained model and thus of the applied first-principles methodology. To compute different observables within the model, we employ ED in the $j_\text{eff}=1/2$ basis on a hexagon-shaped 24-site cluster. We employ all 17 parameters of the main-text Table~I model. Note that restricting to the nearest-neighbor couplings $(J_1,K_1,\Gamma_1,\Gamma'_1)$ of this model gives similar results regarding zigzag order and critical field strengths within ED. But in the following we follow through with all parameters to consistently work with fully \textit{ab-initio} results.

At $\mathbf B = 0 $, $T = 0$, the model correctly reproduces antiferromagnetic zigzag order within ED, revealed by dominant static spin-spin correlations $\braket{\mathbf S(-\mathbf q) \cdot \mathbf S(\mathbf q)}$ at the zigzag ordering wave vectors $\mathbf q \in \{M, M', Y\}$.  
However, significant ferromagnetic correlations $\braket{\mathbf S(0) \cdot \mathbf S(0)} \approx 0.7
\braket{\mathbf S(-M) \cdot \mathbf S(M)}$ are also persistent, as a result of the strong ferromagnetic $|J_1| \approx 0.5 |K_1|$ in the model (Table~I of the main text). This is consistent with the conclusions of a recent study using resonant inelastic X-ray scattering \cite{suzuki2020quantifying}. In the present model, ferromagnetism is so competitive, that the ferromagnetic state ($\mathbf q=0$) is lower in energy than the zigzag one on the classical level. Zigzag order found in ED is therefore likely a result of significant quantum fluctuations. The small $\Gamma'_1=-0.7\,$meV [Table~I in main text] stabilizes this order, and zigzag order is lost for $\Gamma'_1\gtrsim -0.2\,$meV (keeping all other interactions unchanged) within ED. Recalling the effects of compressive \cstar-strain as discussed in the main text, i.e., a strong suppresssion of $|\Gamma'_1|$ and $|\Gamma_1|$ together with a vast increase of $|K_1|$, we estimate compressive uniaxial strains of $\eps \sim -3\%$ to $-5\%$ to be sufficient to suppress zigzag order (at zero magnetic field).

   \begin{figure}
\includegraphics[width=\linewidth]{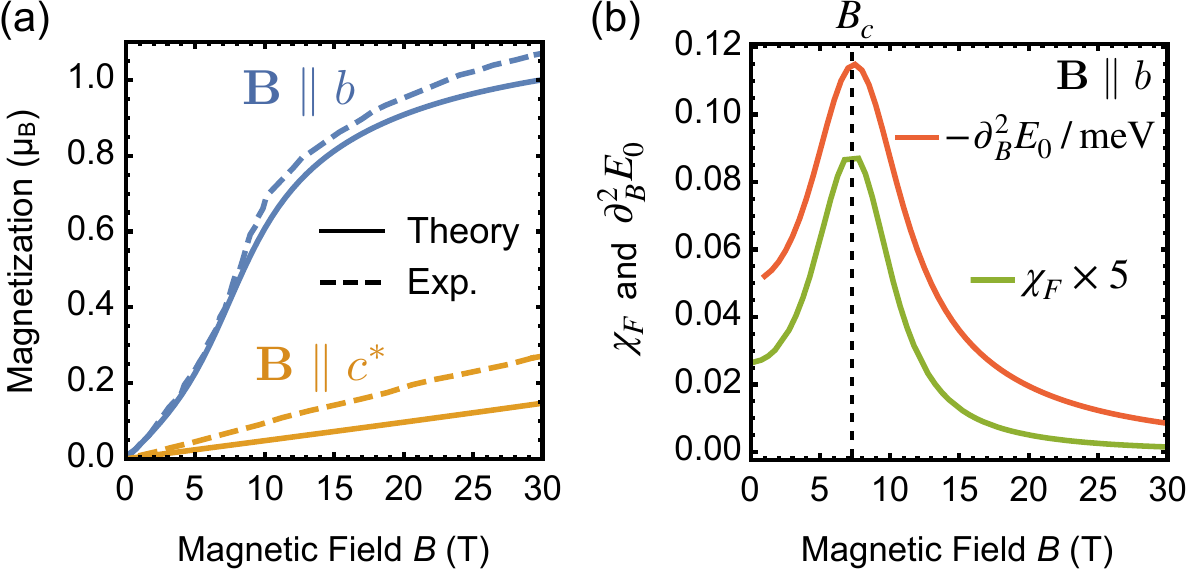}
  \caption{Field-strength-dependent ED results on the model 
  $\{\mathcal J|_{\eps=0}\}$ of Table I in the main text.
  (a)~Magnetization for $\mathbf B\parallel b$ (in-plane) and
  $\mathbf B \parallel \cstar$ (out-of-plane). 
  Dashed curves show experimental data from Ref.~\onlinecite{johnson2015monoclinic}. (b)~Second derivative of the ground-state energy and fidelity susceptibility $\chi_F$. 
  \label{fig:magnetization}
  }
\end{figure}
\begin{figure}
\includegraphics[width=\linewidth]{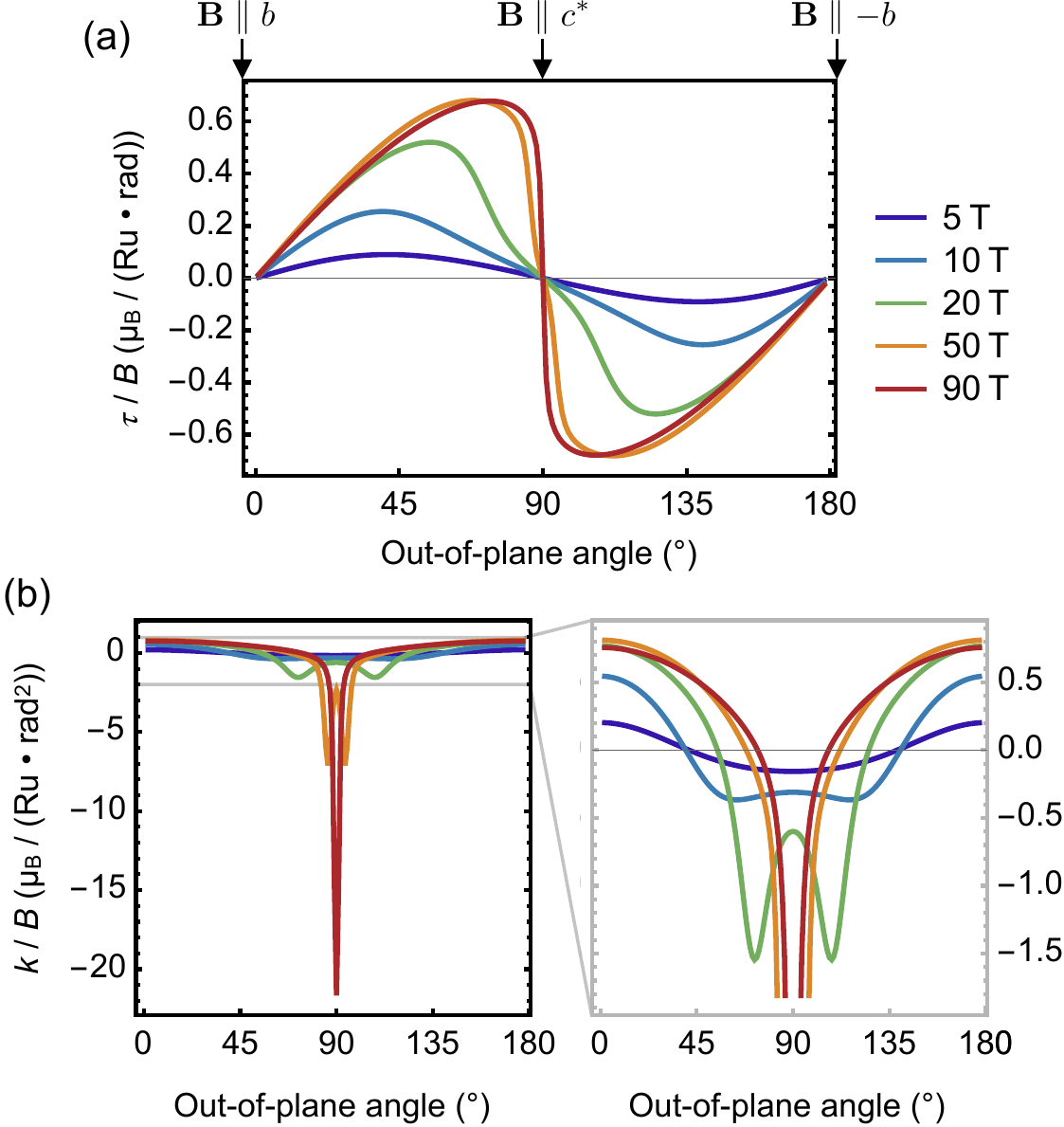}
  \caption{Field-angle-dependent ED results on the model $\{\mathcal J|_{\eps=0}\}$ of Table I in the main text. The field-direction rotates from $b$ (in-plane) over \cstar\ (out-of-plane) to $-b$.  
  (a)~Magnetic torque normalized by field strength, $\tau/B = (dE/d\theta) /B$. (b)~Normalized magnetotropic coefficient $k/B = (d\tau/d\theta) /B$. The lower-right panel shows a narrower plot range  for $k/B$. 
  \label{fig:torque}  }
\end{figure}

We now turn to properties at finite magnetic fields $\mathbf B$. 
\Cref{fig:magnetization}(a) shows the magnetization as a function of field strength for in-plane fields $\mathbf B\parallel b$ and out-of-plane fields $\mathbf B \parallel \cstar$. These are compared to the experimental data of Ref.~\onlinecite{johnson2015monoclinic} at $T<2$\,K. We thus find good agreement. 
To probe for field-induced phase transitions of the model, we show in \cref{fig:magnetization}(b) the fidelity susceptibility $\chi_F=[2/(\delta B)^2][1-\braket{\Psi_0(B)|\Psi_0(B+\delta B)}]$ and the second derivative of the ground state energy ($-\partial^2E_0/\partial B^2$) for in-plane fields $\mathbf B\parallel b$. These reveal a single phase transition as a function of field strength at $B_c\approx 7.5\,$T, between the low-field zigzag ordered phase and the high-field partially-polarized phase. The critical field for the perpendicular direction within the honeycomb plane ($\mathbf B\parallel a$) is found to be similar, while zigzag order is much more stable for fields perpendicular to the plane ($\mathbf B \parallel \cstar$) with $B_c\approx 82\,$T. 
These results are all consistent with experiment, and the physics are analogous to those described in the context of a minimal magnetic model in Ref.~\onlinecite{winte18}.

In \cref{fig:torque}(a) we show the field-angle-dependent magnetic torque $\tau \equiv dE/d\theta$ as a function of the out-of-plane angle $\theta$. 
$\theta=0$ corresponds to the in-plane direction $b$ and $\theta=90^\circ$ to the out-of-plane direction $\cstar$. The essential evolution with field-angle and field-strength and the characteristic sawtooth-shape reproduce the experiments \cite{leahy2017anomalous,modic2018resonant,modic2018chiral} well. \cref{fig:torque}(b) shows the magnetotropic coefficient $k\equiv d\tau/d\theta$, which is also in good qualitative agreement with experiment \cite{modic2018resonant,modic2020scale}. 
The distinct behaviors of $\tau$ and $k$ in the present model are consequences of the significant $\Gamma_1$ interaction and g-anisotropy ($g_{ab}>g_\cstar$) \cite{riedl2019sawtooth}.

\section{Derivation of Equations (3) and (4) of the main text}
We derive Eqs.~(3) and (4) of the main text. These were used to estimate the field-dependent magnetostriction $\lambda_\cstar \equiv L_\cstar^{-1} (\partial L_\cstar/\partial B) = \partial \eps/\partial B$ and the  
 structural Gr\"uneisen
 parameter $\Gruneisen \equiv 
 -\frac{(\partial S_\text{m}/\partial p_\cstar)}{T(\partial S_\text{m}/\partial T)}$. Here, $L_\cstar$ is the length of the crystal along $\cstar$, 
 $S_\text{m}$ the magnetic entropy and $p_\cstar$ uniaxial pressure along \cstar. 
We start with the general expression of the change in the Gibbs free energy under consideration of anisotropic strain and stress contributions
\begin{equation} \label{eq:gibbs}
    \text{d} \mathcal{G}=\sum_{ij} \epsilon_{ij} \,\text{d} \sigma_{ij} -S \,\text{d}T - M \,\text{d}B,
\end{equation}
with the strain tensor $\epsilon_{ij}$, the stress tensor $\sigma_{ij}$ and the crystallographic indices $i,j \in \{a,b,c^\ast\}$. Note that throughout the main text we have used the shorthand notation $\eps_{\cstar \cstar}\equiv \eps$ for uniaxial strain onto $\cstar$. 
\Cref{eq:gibbs} implies a Maxwell relation for the magnetostriction
\begin{equation}
\label{eq:maxwell}
    \lambda_\cstar 
    = \left(\frac{\partial \epsilon_{\cstar\cstar}}{\partial B}\right)_{T,\{\sigma_{ij}\}}
    = \left(\frac{\partial M}{\partial \sigma_{\cstar\cstar}}  \right)_{T,B,\{\sigma_{ij}\}\setminus \{\sigma_{\cstar\cstar}\}}.
\end{equation}
Here, the subscript $\{\sigma_{ij}\}$ denotes that all stress-components are held constant, while $\{\sigma_{ij}\}\setminus \{\sigma_{\cstar\cstar}\}$ holds all stresses 
except $\sigma_{\cstar\cstar}$ constant.

The right-hand side of \cref{eq:maxwell} can also be expressed through the elasticity tensor $c_{ijkl}$, which connects the strain and stress tensors through
\begin{equation}
    \sigma_{ij} = \sum_{kl} c_{ijkl} \, \epsilon_{kl} 
    .
    \label{eq:elasticity}
\end{equation}
Then, we have with \cref{eq:maxwell,eq:elasticity}:
\begin{equation}
\label{eq:elasticity2}
    \left( \frac{\partial M}{\partial \sigma_{c^\ast c^\ast}} \right)_{\heldconstant}
    =  \left[ \sum_{ij} c_{\cstar \cstar ij} \left( \frac{\partial \epsilon_{ij}}{\partial M} \right)_{\heldconstant} \right]^{-1},
\end{equation}
where the variables $\heldconstant=({T,B,\{\sigma_{ij}\}\setminus \{\sigma_{\cstar\cstar}\}})$ are held constant in the derivatives. 

We now employ the assumption that the elasticity contribution along the direction of the investigated length change $\Delta L_\cstar$ is dominant, 
i.e., $c_{\cstar\cstar\cstar\cstar}\gg c_{\cstar\cstar ij}$ for $i,j\neq\cstar$. 
Then we can approximate the magnetostriction as
\begin{equation} \label{eq:lambdadMdc}
    \lambda_{c^\ast} \approx  \frac{\compressibility}{V}  \left( \frac{\partial M}{\partial \epsilon_{\cstar\cstar}} \right)_{\heldconstant},
\end{equation}
where we expressed $c_{\cstar\cstar\cstar\cstar}$ through the linear compressibility under uniaxial pressure $\compressibility= -\frac{\partial \epsilon_{\cstar\cstar}}{\partial p_\cstar} = -V\frac{\partial \epsilon_{\cstar \cstar}}{\partial \sigma_{\cstar \cstar}}=-V c_{\cstar \cstar \cstar \cstar}^{-1}$. 
Note that the derivative in \cref{eq:lambdadMdc} does importantly \emph{not} hold the other lattice constants $a,b$ or the atom positions constant. Instead these degrees of freedom need to be evolved to their new equilibrium positions when $\epsilon_{\cstar\cstar}$ is varied, i.e., the structure needs to be relaxed under constrained variations of $\epsilon_{\cstar\cstar}$, as we have done. 

In \cref{eq:lambdadMdc}, a change in $\epsilon_{\cstar\cstar}$ affects the magnetization $M$ through the variation of the magnetic interactions $\Jij$ and of the $g$-tensor $\mathbb G$ of the pseudospins. This can be expressed formally by applying the chain rule on \cref{eq:lambdadMdc}, which leads to
\begin{equation}
  \lambda_\cstar 
  \approx \frac{\compressibility}{V}  \sum_{{\mathcal J}\in \Jij,\mathbb G} 
  \left[\left( \frac{\partial M}{\partial {\mathcal J}} \right) \left( \frac{\partial {\mathcal J}}{\partial \epsilon_{\cstar\cstar}} \right)\right]_{\heldconstant}. 
  \label{eq:lambdafin}
\end{equation}
Since $\compressibility$ is, up to our knowledge, not known for \rucl, we can not predict the absolute change in $\epsilon_{\cstar\cstar}$ under magnetic field, but instead compute  $\lambda_\cstar / \compressibility$. We thus have to evaluate the derivatives in \cref{eq:lambdafin} at $\epsilon_{\cstar\cstar}=0$. 
While this is an approximation at finite fields, we note that the total integrated field-induced change of $\epsilon_{\cstar\cstar}$ is below $0.1\%$ at $B=15\,$T \cite{gass2020field}, despite \rucl\ having a comparatively large magnetostriction effect. The influence of such small field-induced changes in $\epsilon_{\cstar\cstar}$ onto the field-dependence on the contributions in \cref{eq:lambdafin} is therefore negligible,  cf.~\cref{fig:gvalues,fig:interactions}. 
Instead, the main dependence on magnetic field in \cref{eq:lambdafin} is expected to be carried by the field-dependent derivatives of the magnetization,~$\partial M/\partial \mathcal J$. With the definition $\MEC{\mathcal J}\equiv (\partial \mathcal J/\partial \epsilon_{\cstar\cstar})|_{\epsilon_{\cstar\cstar}=0}$ we thus arrive at 
\begin{equation}
    \lambda_\cstar 
  \approx \frac{\compressibility}{V}  \sum_{{\mathcal J}\in \Jij,\mathbb G} 
  \MEC{\mathcal J}\,\left( \frac{\partial M}{\partial {\mathcal J}} \right)_{\heldconstant,\epsilon_{\cstar\cstar}=0},
  \label{eq:lambdasuppfinal}
\end{equation}
which coincides with Eq.~(3) of the main text. 

For the structural Gr\"uneisen parameter $\Gruneisen\equiv -\frac{\partial S_{\text m}/\partial \sigma_{\cstar\cstar}}{VT(\partial S_{\text m}/\partial T)}$, an analogous approach can be made, where in \cref{eq:elasticity2} $M$ is replaced with $S_\text{m}$. Analogously one arrives at 
\begin{equation}
    \Gruneisen 
   \approx  \frac{\compressibility}{T} \sum_{{\mathcal J} 
     \in{\Jij,\mathbb G}}
 \,  \MEC{\mathcal J} \,
   \left( \frac{\partial S_\text m}{\partial \mathcal J}\right)_{\epsilon_{\cstar\cstar}=0} \left(\frac{\partial S_\text m}{\partial T}\right)^{-1}_{\epsilon_{\cstar\cstar}=0} ,
   \label{eq:suppGruneisenfin}
\end{equation}
which coincides with Eq.~(4) of the main text.

\section{Classical results for magnetostriction}
The results presented in Fig.~4 of the main text are effected by finite-size effects of the 24-site ED calculations. 
Here we compare to a classical 
calculation of the $T=0$ magneto\-striction, that is free of finite-size effects. For this, the magnetization susceptibilities $(\partial M/\partial \mathcal J)$ in \cref{eq:lambdasuppfinal} are evaluated by classical energy minimization of the minimal magnetic model $\{\mathcal J|_{\eps=0}\}$ of Ref.~\onlinecite{winte17} while the magneto\-elastic couplings $\MEC{\mathcal J}$ are taken from Table~I (main text). An analogous classical calculation is not possible on the full \textit{ab-initio}-derived model of Table~I (main text) as that model does not have a zigzag ground state on the classical level, but the insights likely apply also to that model. 

\begin{figure}
    \centering
    \includegraphics[width=0.77\linewidth]{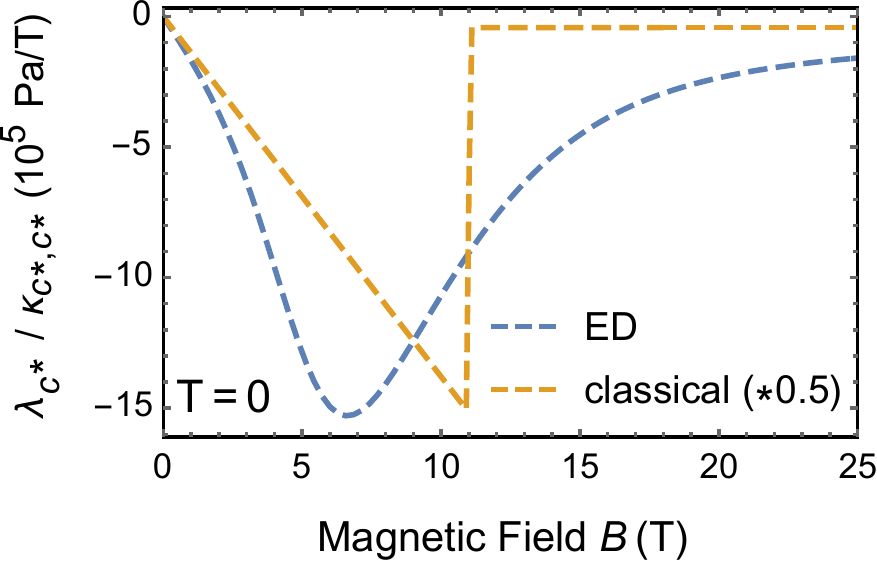}
    \caption{Comparison of ED and classical results for the field-dependent magnetostriction of the minimal magnetic model $\{\mathcal J|_{\eps=0}\}$ of Ref.~\onlinecite{winte17} in conjunction with our magnetoelastic couplings $\MEC{\mathcal J}$ from main-text Table~I. Classical result is divided by a factor of $2$. 
    \label{fig:classical}
    }
\end{figure}

Results are shown in \cref{fig:classical}. Note that the classical result is divided by a factor of $2$  for better comparability. 
Aside from the increased overall magnitude compared to the ED result, 
the drop of the magnitude $|\lambda_\cstar|$ when leaving the zigzag phase is much more pronounced here, which is more consistent with experiment. A smearing-out of phase transitions is expected as a typical finite-size effect in ED. However, the classical result lacks the quantum fluctuations that are present in ED, such that the critical field strength is overestimated ($B_{c,\text{classical}}=11\,$T, $B_{c,\text{ED}}=6\,$T). 

Both in the classical and in the ED results, we note that the dominating contribution to $\lambda_\cstar$ comes from the summand in \cref{eq:lambdasuppfinal} that describes magnetoelastic coupling from  $\MEC{\Gamma'_1}$.  
An increase in $\Gamma'_1$ lowers the in-plane critical field strength \cite{maksimov2020rethink}, such that ---within the zigzag phase $B<B_c$--- spins are more strongly canted towards the direction of the magnetic field for a given field strength $B$, i.e., have higher magnetization. Therefore $(\partial M/\partial \Gamma'_1)>0$, 
which together with $\widetilde{\Gamma'_1}<0$ [Table~I (main text)] explains an overall \emph{negative} sign of the magnetostriction.

\section{Details on structural Gr\"uneisen parameter}
\begin{figure}
    \centering
    \includegraphics[width=0.9\linewidth]{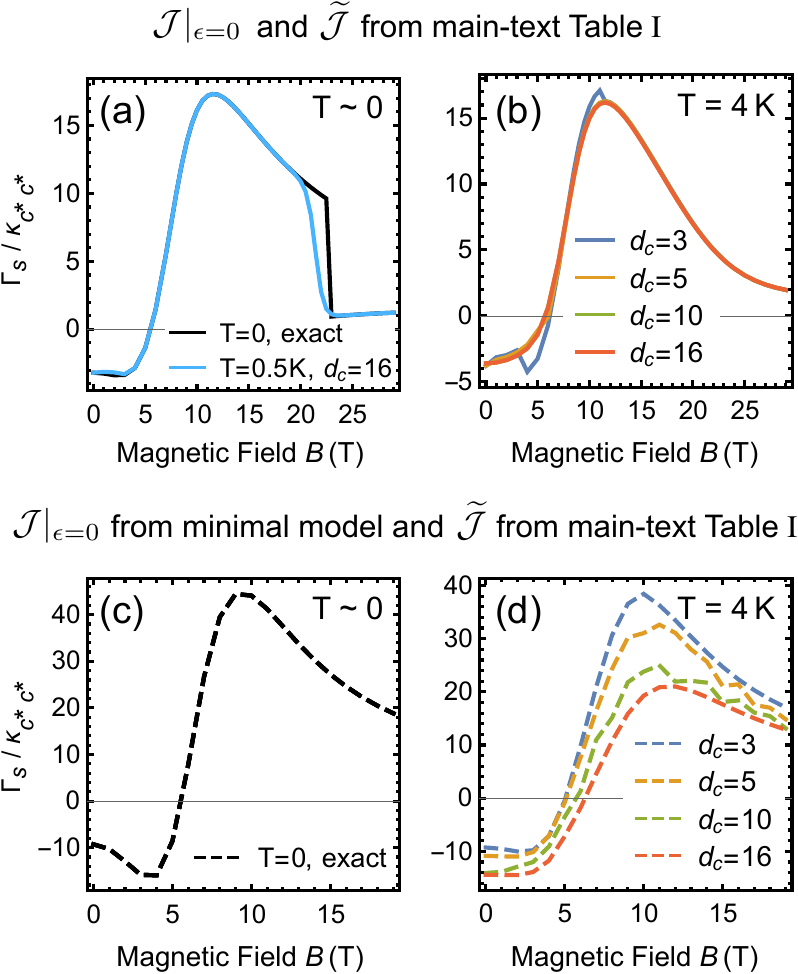}
    \caption{
    Detailed results on temperature-dependence and estimation of the cutoff errors for the structural Gr\"uneisen parameter \Gruneisen. Top panels (a,b) show results for $\mathcal J|_{\eps=0}$ and $\MEC{\mathcal J}$ from main-text Table I, bottom panels (c,d) show results for $J|_{\eps=0}$ from the minimal magnetic model of Ref.~\onlinecite{winte17} and $\MEC{\mathcal J}$ from main-text Table I. (a,c)~Zero- and low-temperature results. (b,d)
   ~Results at $T=4\,$K for different employed cutoffs $d_c$ (see \cref{eq:cutoff}). 
    \label{fig:finiteT}
    }
\end{figure}

The structural Gr\"uneisen parameter $\Gruneisen \equiv 
 -\frac{(\partial S_\text{m}/\partial p_\cstar)}{T(\partial S_\text{m}/\partial T)}
 $ can not be obtained from a pure ground state calculation (where the entropy would vanish). As we are interested in low-but-finite temperature calculations, we employ a method where we restrict the canonical sums to the lowest $d_c$ eigenstates,
  \begin{align}
  Z&\approx \sum_{n=0}^{d_c-1} e^{-E_n/(k_BT)},\nonumber
  \\  \braket{O}&\approx \frac1Z  \sum_{n=0}^{d_c-1} e^{- E_n/(k_B T)} \braket{n|O|n},
  \label{eq:cutoff}
\end{align}
which has proven reliable for calculations of the magnetic Gr\"uneisen parameter $\Gamma_B$ \cite{bachus2020thermodynamic}. We use \cref{eq:cutoff} to evaluate the structural Gr\"uneisen parameter at $T=4\,$K via \cref{eq:suppGruneisenfin} and ED with up to $d_c=16$. The dependence on $d_c$ of our results is shown in \cref{fig:finiteT}(b), which corresponds for $d_c=16$ to the solid line in Fig.~4(c) of the main text.  As these results are very robust already for $d_c>3$, we deem the results very reliable. 
In \cref{fig:finiteT}(d) we show the error analysis for the case where the minimal magnetic model of Ref.~\onlinecite{winte17} is used for the $\mathcal J|_{\eps=0}$ couplings. Here, results appear not fully converged at $T=4\,$K, but we assume trends to be correct. This can also be seen at the exact $T=0$ result shown in \cref{fig:finiteT}(c) of the same model. Here an exact evaluation is possible as the $T\rightarrow 0$ limit of a Gr\"uneisen parameter is only determined by the behavior of the gap between the ground state and lowest excited state  \cite{bachus2020thermodynamic}. 

Inspecting the zero-temperature limit of $\Gruneisen$ in our 
full \textit{ab-initio} model from the main-text Table~I, shown in
\cref{fig:finiteT}(a), a discontinuity is apparent at $B=22.5\,$T. 
This is the result of a level crossing in the first two excited states at this field strength, which produces such an anomaly in all Gr\"uneisen parameters of the form  $\Gamma_\lambda\equiv - \frac{\partial S/\partial \lambda}{T(\partial S/\partial T)}$ \cite{bachus2020thermodynamic}. 
At small finite temperatures [blue curve in \cref{fig:finiteT}(a)], the discontinuity is smeared out to a shoulderlike feature, and is mostly invisible for intermediate and high temperatures, as in \cref{fig:finiteT}(b) for $T=4$\,K. 
An analogous shoulder-anomaly has been observed experimentally in the \textit{magnetic} Gr\"uneisen parameter $\Gamma_B$ at $B\sim 10\,$T, that resembles the anomaly in $\Gamma_B$ in our model \cite{bachus2020thermodynamic}. 
While the location of the shoulder-anomalies in this model and experiment appear far apart in field strength, we note that ---up to our knowledge--- no other realistic model proposed so far for \rucl\ (including the minimal model \cite{winte17} we discussed) features any such shoulder-anomaly in Gr\"uneisen parameters. In the present model, a level crossing between the lowest excited states, that produces the shoulder-anomaly, happens between states at $\mathbf k = Y$ and $\mathbf k=0$. We interpret this to be a result of the strongly competing ferromagnetic phase (with ordering wave vector $\mathbf{q}=0$) in the present model. As the precise field strength at which the anomaly occurs is highly sensitive to the coupling strengths, future refinements of the model may take this into account. 

We note that  the present full \textit{ab-initio} model predicts the low-temperature shoulder anomaly to be significantly stronger in the structural Gr\"uneisen parameter $\Gruneisen$ [\cref{fig:finiteT}(a)] than in the magnetic Gr\"uneisen parameter $\Gamma_B$ \cite{bachus2020thermodynamic}. Since the field strength at which the anomaly takes place is overestimated in this model, the actual drop in $\Gruneisen$ might occur already at $B\approx 10\,T$ \cite{bachus2020thermodynamic}. Accordingly, a small dip in  $\Gruneisen$ is suggested by experimental data at $T=3.5\,$K \cite{gass2020field}.

\section{Further Details of First-principles Calculations}

\itsection{Structural relaxation}
The constrained relaxations were performed using {\it ab-initio} DFT as implemented in the QUANTUM ESPRESSO (QE) package~\cite{qe} in zigzag antiferromagnetic configurations of ruthenium. A plane-wave basis set was used to expand the electronic wave functions and the exchange-correlation functional was approximated by the generalized gradient approximation (GGA+$U$) of Perdew, Burke, and Ernzerhof~\cite{pbe} with $U=1.5$\,eV. 
 The cutoff for the plane-wave basis set and the cutoff for the corresponding charge densities were set at 60 and 600\,Ry, respectively. 
 We considered Van-der-Waals corrections within Grimme's DFT-D2 method~\cite{grimme}. A Monkhorst-Pack~\cite{mp} grid of size 8$\times$6$\times$8 was used to generate the k-mesh (zone centered) for the corresponding Brillouin-zone sampling.

In order to check the effects of spin-orbit coupling (SOC) on the relaxation, we used VASP code~\cite{vasp} using the projector-augmented planewave basis~\cite{blochl}. For this, the unstrained structural optimization at ambient pressure was recalculated with  VASP in the GGA+$U$ approximation and the results from the two methods were found to agree well. Then the former code was used to find the effect of SOC (within GGA+$U$+SOC) on the lattice  geometry. 
We find that SOC mainly brings the structure further to approximate $C_3$ symmetry of the honeycomb planes. This effect is in line with other studies of honeycomb ruthenates, iridates and rhodates~\cite{kim2016crystal,hermann2018competition,hermann2019pressure} and consistent with the approximately $C_3$-symmetric magnetic response observed in \mbox{\rucl} \cite{johnson2015monoclinic,banerjee2017neutron,do2017majorana,lampen2018field}. 
These observations also vindicate our work with a $C_3$-simplified version of the obtained model throughout the discussions in the main text. 
In \cref{tab:couplingsfull} we give the full model before $C_3$ symmetrization, where coupling strengths on Z bonds can differ from those on X/Y bonds. The rows $\mathcal J_\mathrm{3orb}$ and $\MEC{\mathcal J_\mathrm{3orb}}$ in this table show results not including effects of all 5 orbitals, that are compared below.

\itsection{Exchange interaction calculations}
With the structural response toward uniaxial strain $\eps$ established, each structure can be associated with a magnetic Hamiltonian determined by $\Jij$ and $\mathbb G$ (see Eq.~(1) of the main text). To extract the corresponding coupling parameters $\mathcal J \in \Jij$, we first construct the multi-orbital Hubbard Hamiltonian for the ruthenium sites.

For the two-particle interaction terms, we used spherically symmetric expressions~\cite{slater1960quantum}, parametrized by Slater integrals, and fixing them to values based on recent cRPA results for \rucl~\cite{eichstaedt2019deriving}. The local two-particle exchange parameters were obtained by subtraction of the non-local contributions given in Ref.~\onlinecite{eichstaedt2019deriving}. Averaging over the orbital-dependent expression and taking an effective two-particle on-site interaction then lead to the parameters employed in this work, 
$U_{t_{2g}} = F_0 + (4/49)(F_2 + F_4) = 1.68\,$eV 
 and $J_{t_{2g}} = (3/49)F_2 + (20/441)F_4 = 0.29\,$eV.

\begin{figure}
\includegraphics[width=\linewidth]{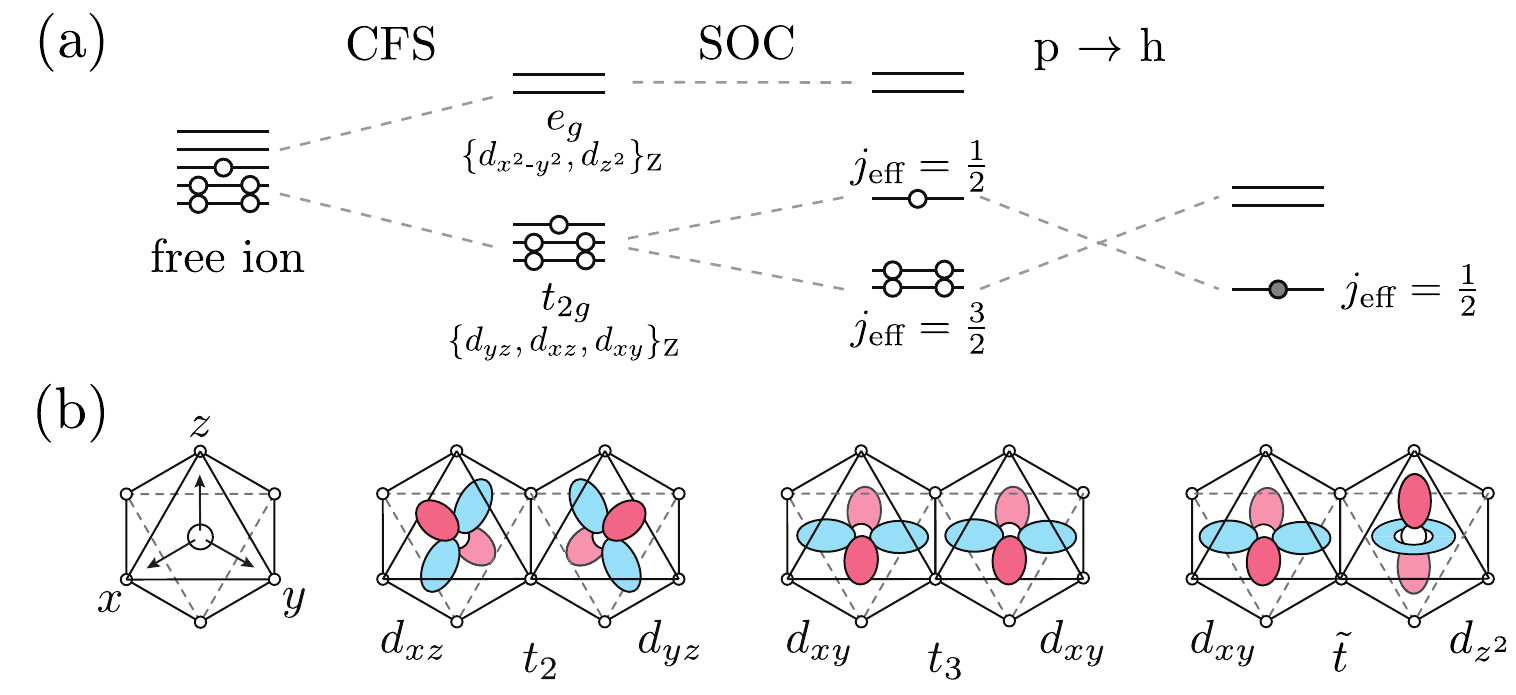}
  \caption{(a) Effects of crystal field splitting (CFS), spin-orbit coupling (SOC) and a particle-hole ($p \rightarrow h$) transformation in a d$^5$ configuration. (b) Dominant hopping processes $t_2$, $t_3$ and $\tilde{t}$ on the Z-bond for the experimental structure of \rucl, explicitly given in Eq.~\eqref{eq:exp_hop}. \label{fig:hoppings} }
\end{figure}

The material-specific properties are encoded in the first-principles ruthenium hopping parameters. As illustrated in Fig.~\ref{fig:hoppings}(a), the crystal field splitting (CFS) due to the octahedral chlorine environment causes a splitting between energetically low $t_{2g}$ orbitals and the higher in energy $e_g$ orbitals. Due to the energy gap the low-energy properties of the system can be described in terms of a low-spin configuration, where the electrons populate the $t_{2g}$ orbitals.
Spin-orbit coupling (SOC) was included in the atomic approximation with a SOC parameter $\lambda=0.15\,$eV~\cite{montalti2006handbook}. 
Considering such strong spin-orbit coupling, the $t_{2g}$ orbitals form together with the spin 1/2 degree of freedom $j_{\text{eff}}=1/2$ and $j_{\text{eff}}=3/2$ states [see \cref{fig:hoppings}(a)]. Since the $j_{\text{eff}}=3/2$ states are fully occupied, exact diagonalization of the two-site five-orbital Hubbard Hamiltonian then allows to project the determined low-energy states onto a bilinear pseudospin 1/2 Hamiltonian. 
The resulting bond-resolved first-principles coupling parameters up to third-nearest neighbors at $\eps=0$ are shown in \cref{tab:couplingsfull}. $C_3$-symmetrization of these values lead to the parameters given in the main text.

\begin{table*}
\begin{tabular}{L||C|C|C||C|C|C|C||C|C|C|C|C|C|C||C|C|C|C}
\hline
 &g_a & g_b  & g_\cstar & J_1 & K_1 & \Gamma_1 & \Gamma'_1 & J_2 & K_2 & \Gamma_2 & \Gamma'_2 & D^\alpha_{2} & D^\beta_{2}& D^\gamma_{2} & J_3 & K_3 & \Gamma_3 & \Gamma'_3  \\
 \hline
    \mathcal J^{\text{Z}}_{5\text{orb}}
  & & & & {-5.}&	{-12.4}&	{6.9}&	{-1.2}&	0.&		-0.1&	0.&	-0.&	0.&	0.&	0.2&		0.3& 0.2& 0.1& -0.1 \\
    \mathcal J^{\text{X}}_{5\text{orb}}
  & \multirow{-2}{*}{${2.27}$} & \multirow{-2}{*}{${2.44}$} & \multirow{-2}{*}{${1.88}$} &
  {-6.}&	{-9.}&	{10.6}&	{-0.5}&	0.&		-0.2&	0.1&	0.1&	0.&	0.1&	0.&		0.2& 0.3& 0.2& -0.1 \\ 
  \hline
      \mathcal J^{\text{Z}}_{3\text{orb}}
  & & & & {-2.4}&	{-7.7}&	{5.5}&	{-1.4}&	-0.3&		-0.6&	0.1&	-0.2&	-0.&	-0.&	0.&		0.3& 0.3& -0.1& -0.1 \\
   \mathcal J^{\text{X}}_{3\text{orb}}
  & \multirow{-2}{*}{${2.07}$} & \multirow{-2}{*}{${2.25}$} & \multirow{-2}{*}{${1.50}$} &
  {-4.7}&	{-2.9}&	{10.1}&	{-0.9}&	-0.8&		0.5&	-0.1&	0.1&	-0.1&	0.1&	-0.1&		0.1& 0.3& -0.2& -0.2 \\ 
  \hline \hline
    \MEC{\mathcal J^{\text{Z}}_{5\text{orb}}}
  & & & & {-15.8}&	{35.}&	{3.4}&	{-9.1}&	-1.4&		1.9&	-0.8&	-0.4&	-1.&	-1.&	-3.9&		3.8& 1.& -0.4& -0.5 \\
    \MEC{\mathcal J^{\text{X}}_{5\text{orb}}}
  & \multirow{-2}{*}{${-1.38}$} & \multirow{-2}{*}{${-1.84}$} & \multirow{-2}{*}{${3.85}$} &
  {9.9}&	{43.3}&	{9.5}&	{-12.7}&	-0.6&		1.4&	-0.3&	0.1&	-1.3&	-0.8&	-2.8&		0.6& 0.3& -0.7& -0.5 \\
  \hline
    \MEC{\mathcal J^{\text{Z}}_{3\text{orb}}}
  & & & & {-9.4}&	{43.3}&	{8.}&	{-1.8}&	0.4&		2.4&	-1.4&	-0.8&	-1.&	-1.&	-2.8&		3.5& 1.6& -1.3& -0.8 \\
    \MEC{\mathcal J^{\text{X}}_{3\text{orb}}}
  & \multirow{-2}{*}{-2.08} & \multirow{-2}{*}{-2.43} & \multirow{-2}{*}{5.07} &
  {-14.3}&	{61.}&	{17.8}&	{-7.5}&	-0.4&		0.6&	-0.6&	-0.2& 	-1.4&	-0.4&	-1.9&		0.1& 0.4& -1.& -0.7 \\
  \hline
\end{tabular}
\caption{
$g$-tensor components and magnetic interactions $\mathcal J^\text Z|_{\eps=0}$ ($\mathcal J^\text X|_{\eps=0}$) on Z-bonds (X-bonds) in meV for the zero-strain relaxed $C2/m$ structure before $C_3$ symmetrization and with hopping parameters considering all five Ru $4d$ orbitals ($\mathcal{J}_\text{5orb}$) compared to only Ru $t_{2g}$ orbitals ($\mathcal{J}_\text{3orb}$). Y bonds are related to X bonds by mirror symmetry. 
\label{tab:couplingsfull}}
\end{table*}

 \begin{figure}
\includegraphics[width=\linewidth]{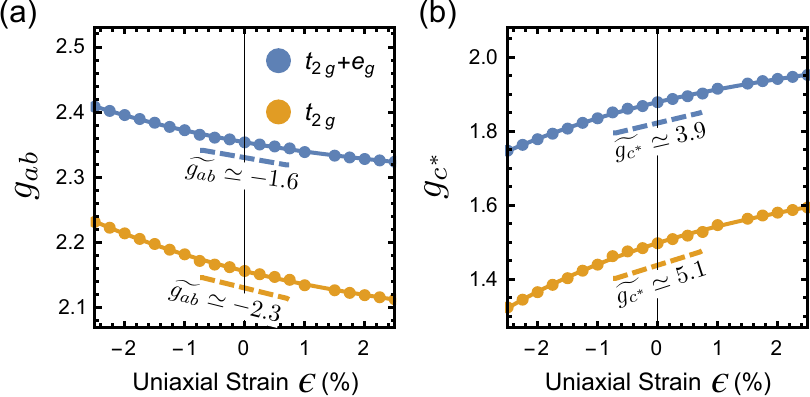}
  \caption{$g$-values at different uniaxial strains \eps, either taking into account only the $t_{2g}$ orbitals [CAS(5,3)] or all five $d$-orbitals [$t_{2g}+e_g$; CAS(5,5)]. Solid lines show third-order polynomial fits. Values next to  dashed lines indicate $\MEC{\mathcal J}\equiv (\partial J/\partial \eps) |_{\eps=0}$. 
   \label{fig:gvalues} 
   }
     \vspace{\floatsep}
\includegraphics[width=\linewidth]{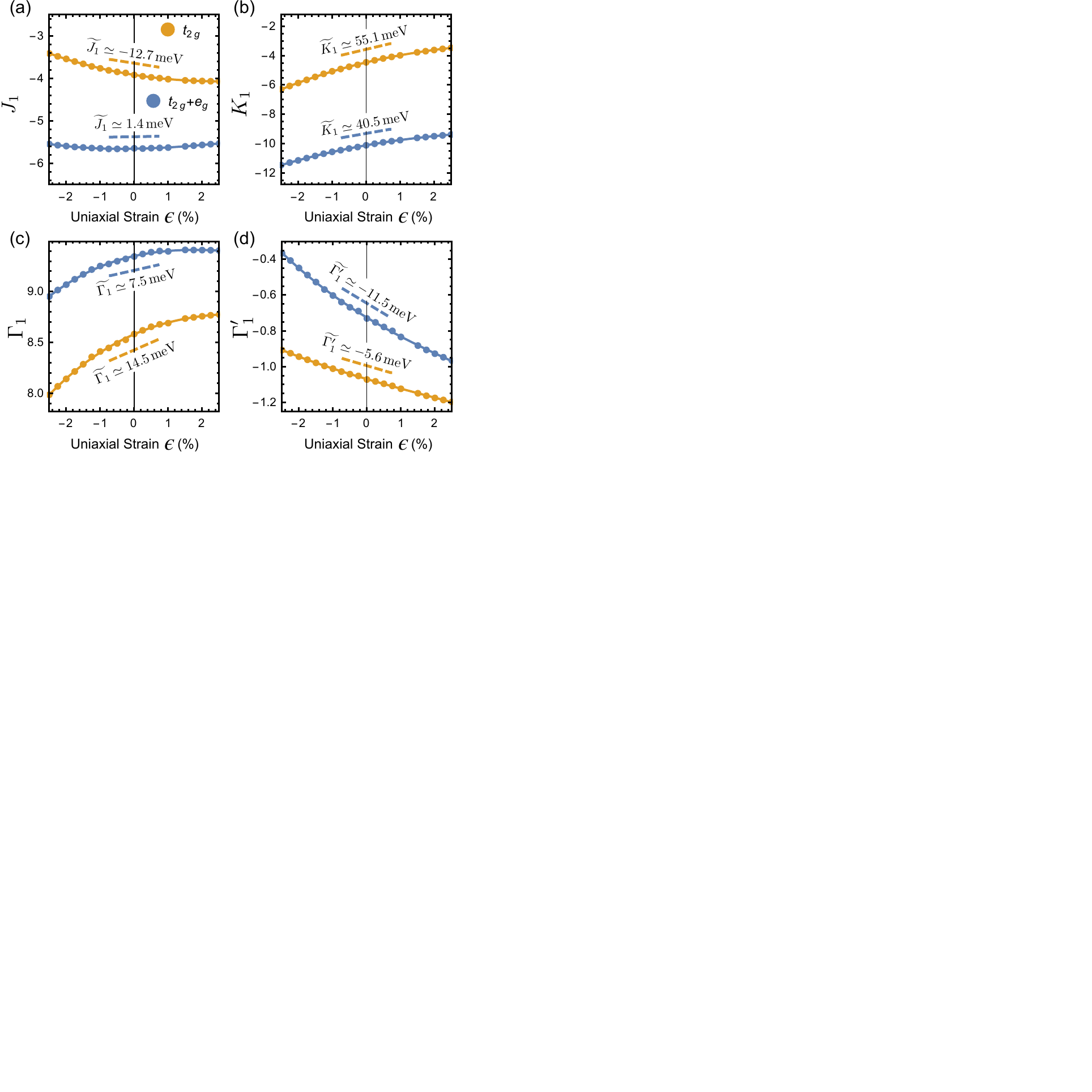}
  \caption{Nearest-neighbor magnetic interactions at different uniaxial strains \eps, either taking into account only the $t_{2g}$ orbitals or all five $d$-orbitals ($t_{2g}+e_g$). Solid lines show third-order polynomial fits. Values next to dashed lines indicate $\MEC{\mathcal J}\equiv (\partial J/\partial \eps) |_{\eps=0}$. 
  \label{fig:interactions}
   }
\end{figure}

\section{Effects of $e_g$ orbitals}

Although the nearest-neighbor Kitaev interaction was introduced within the three-orbital $t_{2g}$ framework of the Khaliullin-Jackeli mechanism~\cite{jackeli2009mott,rau2014trigonal,winter2017models}, the role of $e_g$ orbitals for Kitaev materials was previously considered in some approximations~\cite{chaloupka2013zigzag,foyevtsova2013abinitio,kim2015kitaev}.
In this work, we extended the approach by some of the authors~\cite{winte16}, where the Hubbard Hamiltonian was constructed with only $t_{2g}$ ruthenium orbitals. 
Interestingly, inclusion of the $e_g$ orbitals leads to an increase of the magnitudes of the $g$ tensor components as well as the bilinear magnetic interactions, given in Table~\ref{tab:couplingsfull}, including the ferromagnetic nearest-neighbor Kitaev interaction $K_1$. Meanwhile, the magnetoelastic couplings are in most cases overestimated if $e_g$ effects are neglected. 

If only $t_{2g}$ orbitals were considered,
the relevant Hilbert space could be reduced by a particle-hole transformation, which results in a projection of the one-hole low-energy solution onto pseudospin 1/2 operators [see \cref{fig:hoppings}(a)].
Additional consideration of $e_g$ orbitals within this procedure takes higher-order hopping processes into account that lead to corrections of the final effective pseudo\-spin Hamiltonian.
The present analysis is limited to two-site clusters, which likely leads to underestimation of second and third neighbor couplings, which were found to be of non-negligible magnitude~\cite{winte16,hou2017unveiling}. However, this restriction is necessary to limit computational expense.

In order to highlight the effects of the $e_g$ orbitals on the $g$-tensor, we compare the results obtained from ORCA \cite{neese2012orca} at the CASSCF/TPSSh/def2-TZVP level on [RuCl$_6]^{3-}$ clusters, using active space definitions of (5,3) and (5,5). The former explicitly excludes any configurations with partial occupancy of the $e_g$ orbitals, while the latter includes those configurations. In both cases, we considered equal weight on all doublet states in the orbital optimization. The $e_g$ effects for the  $C_3$-symmetrized $g$ tensor components as a function of uniaxial strain $\eps$ are illustrated in \cref{fig:gvalues}. At $\eps=0$, the full approach considering $(t_{2g}+e_g)$ orbitals reveals an increased magnitude for both, $g_{ab}$ and $g_\cstar$. This increase is not homogeneous, leading to a reduction of the anisotropy $g_{ab}/g_\cstar$. Considering the non-symmetrized values in \cref{tab:couplingsfull}, it also becomes evident that the anisotropy within the honeycomb plane, i.e., $g_a<g_b$, is reduced by $e_g$ effects.
The magnetoelastic couplings $\MEC{g_{ab}}$ and $\MEC{g_\cstar}$ are reduced upon consideration of these effects, so that an overestimation of the coupling to the lattice can be prevented by consideration of such higher-order processes.

In \cref{fig:interactions} we compare strain-dependent nearest-neighbor magnetic interactions considering $t_{2g}$ and $(t_{2g}+e_g)$ orbitals. 
In this case, $e_g$ effects can be related to the first-principles hopping parameters between them and the low-energy $t_{2g}$ orbitals. For the $\eps=0$ structure we computed the following parameters on the Z-bond (the bond parallel to the $b$ direction, see Fig.~1 of the main text):
\begin{align} \label{eq:exp_hop}
t_Z = \left(
 \begin{array}{c|ccccc} 
 & d_{yz}& d_{xz}& d_{xy}& 	d_{x^2\text{-}y^2}& d_{z^2} \\
 \hline
 d_{yz} & +0.04& \mathbf{+0.15}& -0.01& 0 & 0 \\  
 d_{xz} & \mathbf{+0.15}& +0.04& -0.01& 0& 0 \\ 
 d_{xy} & -0.01& -0.01& \mathbf{-0.08}& 0& \mathbf{+0.28} \\
 d_{x^2\text{-}y^2}& 0& 0& 0& -0.01& 0 \\ 
 d_{z^2} & 0& 0& \mathbf{+0.28}& 0& 0
 \end{array}\right)
\end{align}
The highlighted dominant hopping mechanisms are illustrated in Fig.~\ref{fig:hoppings}(b). While $t_2$ and $t_3$ are strong hoppings within the $t_{2g}$ orbitals, we find the strongest hopping to be $\tilde{t}$, which connects the $t_{2g}$ orbital $d_{xy}$ with the $e_g$ orbital $d_{z^2}$. 

Consideration of this additional large exchange mechanism leads to a reshuffling of relative coupling strengths with an overall tendency to an increase in magnitude (see also Table~\ref{tab:couplingsfull}). 
For $K_1$, $\Gamma_1$, and $J_1$, the absolute value increases including $e_g$ orbitals.
The only reduced nearest-neighbor interaction is $\Gamma_1^\prime$.

The magnetoelastic couplings $\MEC{\mathcal{J}}$ are given in Table~\ref{tab:couplingsfull} and are illustrated by the slope of the dashed lines in \cref{fig:interactions}.  For most interaction parameters they follow the same trend when including $e_g$ effects, but become reduced in magnitude. Exceptions are $\MEC{\Gamma'_1}$ and $\MEC{J_1}$: 
For $\MEC{\Gamma'_1}$ the inclusion of $e_g$ effects leads to a strongly \emph{enhanced} magnetoelastic coupling compared to the 3-orbital result. 
For $\MEC{J_1}$, the five-orbital result shows differing signs on Z and X/Y bonds. 
This leads to a (possibly artificially) small value of the $C_3$-symmetrized $\MEC{J_1}=(2\MEC{J_1^\text{X}}+\MEC{J_1^\text{Z}})/3$. 
However, as the experimentally accessible magnetostriction and structural Gr\"uneisen parameter are dominated by the $\MEC{\Gamma_1^\prime}$ response, this issue has little consequence for the quantities discussed in this work and we therefore decided to keep the discussion in the $C_3$-symmetrized limit. 

We therefore conclude that the interplay of higher order hopping processes together with the structure of 
Hund's coupling in $d$ block elements lead to a delicate coupling between magnetism and structure that ---to the best of
our knowledge—-- has not been captured with analytical
methods such as perturbation theory so far.

 \section{Inter-plane couplings}
 In \rucl, most phenomena have generally been well captured qualitatively in terms of quasi-two-dimensional descriptions, that neglect inter-plane magnetic couplings between the van-der-Waals layers \cite{rau2016spin,winter2017models,laurell2020dynamical}. 
 As uniaxial strain along \cstar\ however affects the distance between the honeycomb planes stronger than in-plane distances (see Fig.~2 of the main text), one might suspect \emph{magnetoelastic} couplings related to inter-plane couplings to become significant. 
 
 \begin{figure}
\includegraphics[width=0.9\linewidth]{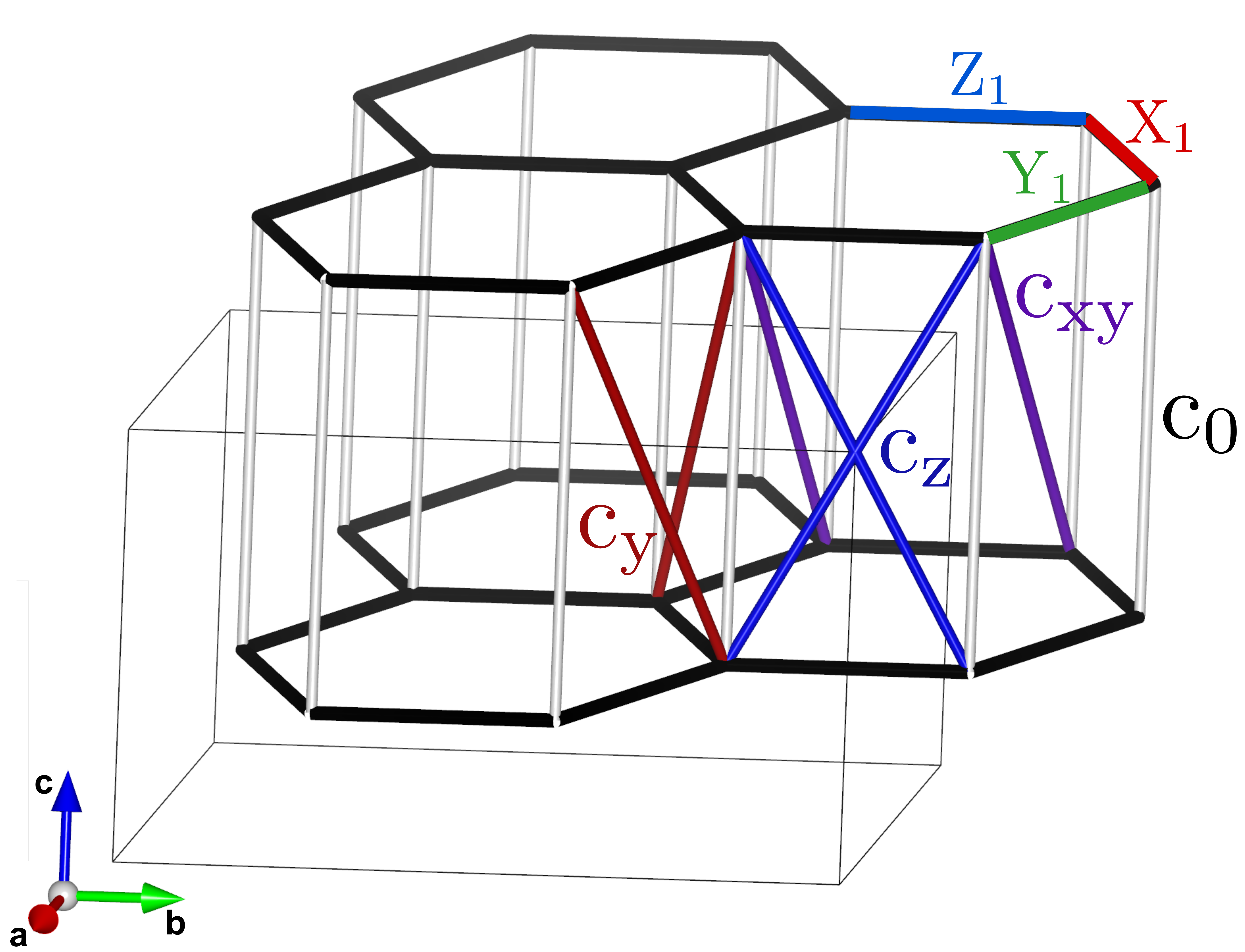}
  \caption{
  Four shortest inter-layer bonds in the relaxed $C2/m$ \rucl structure at ambient pressure: c$_0$ connects two ruthenium sites by the crystallographic $c$ axis, c$_\text{y}$ is the bond of two sites that are connected by the inter-layer bond c$_0$ and the intra-layer bond Y$_1$, c$_\text{z}$ connects sites along c$_0$--Z$_1$, and c$_\text{xy}$ connects sites along c$_0$--X$_1$--Y$_1$.
   \label{fig:inter-layer} 
   }
\end{figure}

 We have therefore calculated the magnetoelastic couplings of the four shortest inter-layer bonds, illustrated in Fig.~\ref{fig:inter-layer}, with the same procedures as described above for the in-plane couplings.
For the shortest-distance inter-plane bond, which connects two ruthenium sites by the crystallographic $c$ axis, labeled ``$c_0$'', we extract a magnetic pseudospin interaction $H_{ij}=\mathbf S_i \cdot \mathbb J_{ij} \cdot \mathbf S_j$ with 
  \begin{align}
     \mathbb J_{\text{c}_0} = 
     \left(
 \begin{array}{c|ccc} 
 &x&y&z\\
 \hline
 x & 0.06 & 0.02 & 0.05 \\
 y & 0.07 & 0.06 & 0.05 \\
 z & 0.07 & 0.07 & -0.06 \\
\end{array}
\right) \, \text{meV}, 
\end{align}
 which is much weaker than the nearest-neighbor in-plane magnetic exchange ($K_1=-10.1\,$meV). For the corresponding magnetoelastic couplings on the same inter-plane bond we extract
 \begin{align}
 \MEC{\mathbb J_{\text{c}_0}} &= 
\left(
 \begin{array}{c|ccc} 
 &x&y&z\\
 \hline
 x &-0.67 & 0.47 & -0.37 \\
 y &-0.63 & -0.67 & -0.37 \\
 z &-0.86 & -0.86 & 0.4 \\
\end{array}
\right)
\, \text{meV} 
 \end{align}
 which should be compared to the in-plane $\MEC{K_1}=40.5\,$meV and $\MEC{\Gamma'_1}=-11.5\,$meV.

We labelled the next-shortest inter-layer bond ``c$_\text{y}$'', which is the bond of two sites that are connected by the inter-layer bond c$_0$ and the intra-layer bond Y$_1$ (see Fig.~\ref{fig:inter-layer}). We find the couplings to be of similar order of magnitude:
   \begin{align}
     \mathbb J_{\text{c}_\text{y}} = 
     \left(
 \begin{array}{c|ccc} 
 &x&y&z\\
 \hline
 x & -0.09 & 0.07 & 0.02 \\
 y & 0.07 & 0.11 & 0.01 \\
 z & 0.02 & 0.01 & 0.09 \\
\end{array}
\right) \, \text{meV}. 
\end{align}
Interestingly, the corresponding magnetoelastic couplings are slightly increased compared to the values for bond c$_0$:
 \begin{align}
 \MEC{\mathbb J_{\text{c}_\text{y}}} &= 
\left(
 \begin{array}{c|ccc} 
 &x&y&z\\
 \hline
 x & 0.04 & -0.17 & -0.28 \\
 y & -0.17 & -1.24 & 0.27 \\
 z &-0.28 & 0.27 & -0.86 \\
\end{array}
\right)
\, \text{meV} 
 \end{align}
The same is true for the c$_\text{z}$ bond that connects sites along the path c$_0$--Z$_1$:
   \begin{align}
     \mathbb J_{\text{c}_\text{z}} = 
     \left(
 \begin{array}{c|ccc} 
 &x&y&z\\
 \hline
 x & 0.08 & -0.01 & -0.03 \\
 y & -0.01 & 0.2 & -0.03 \\
 z & -0.03 & -0.03 & 0.03 \\
\end{array}
\right) \, \text{meV}, 
\end{align}
with
 \begin{align}
 \MEC{\mathbb J_{\text{c}_\text{z}}} &= 
 \left(
 \begin{array}{c|ccc} 
 &x&y&z\\
 \hline
 x & -1.59 & 0. & 0.06 \\
 y & 0. & -3.19 & 0.06 \\
 z & 0.06 & 0.06 & -0.89 \\
\end{array}
\right)
\, \text{meV} .
 \end{align}
Finally, two sites connected by the path c$_0$--X$_1$--Y$_1$ contain slightly smaller couplings and magnetoelastic couplings:
   \begin{align}
     \mathbb J_{\text{c}_\text{xy}} = 
     \left(
 \begin{array}{c|ccc} 
 &x&y&z\\
 \hline
 x & 0.05 & -0.06 & 0.08 \\
 y & 0.1 & 0.05 & 0.08 \\
 z & -0.13 & -0.13 & 0.17 \\
\end{array}
\right) \, \text{meV}, 
\end{align}
with
 \begin{align}
 \MEC{\mathbb J_{\text{c}_\text{xy}}} &= 
\left(
 \begin{array}{c|ccc} 
 &x&y&z\\
 \hline
 x &  0.29 & -0.48 & -0.63 \\
 y & -0.63 & 0.29 & -0.63 \\
 z & 0.74 & 0.74 & -0.46 \\
\end{array}
\right)
\, \text{meV} 
 \end{align}
 
 Inter-plane couplings have therefore been neglected in the discussion of magnetostriction and structural Gr\"uneisen parameter.

\end{document}